\documentclass[preprint2]{aastex}
\usepackage{rotating}
\usepackage{graphicx}
\usepackage{grffile}

\shortauthors{Crepp \& Johnson 2011}

\begin{document}
\setcounter{secnumdepth}{3}
\title{Estimates of the Planet Yield from Ground-Based High-Contrast Imaging Observations as a Function of Stellar Mass}

\author{Justin R. Crepp \& John Asher Johnson}
\affil{California Institute of Technology, 1200 E. California Blvd., Pasadena, CA 91125} 
\email{jcrepp@astro.caltech.edu} 

\begin{abstract}  
We use Monte Carlo simulations to estimate the number of extrasolar planets that are directly detectable in the solar-neighborhood using current and forthcoming high-contrast imaging instruments. Our calculations take into consideration the important factors that govern the likelihood for imaging a planet, including the statistical properties of stars in the solar neighborhood, correlations between star and planet properties, observational effects, and selection criterion. We consider several different ground-based surveys, both biased and unbiased, and express the resulting planet yields as a function of stellar mass. Selecting targets based on their youth and visual brightness, we find that strong correlations between star mass and planet properties are required to reproduce high-contrast imaging results to date. Using the most recent empirical findings for the occurrence rate of gas-giant planets from RV surveys, our simulations indicate that naive extrapolation of the Doppler planet population to separations accessible to high-contrast instruments provides an excellent agreement between simulations and observations using present-day contrast levels. In addition to being intrinsically young and sufficiently bright to serve as their own beacon for adaptive optics correction, A-stars have a high planet occurrence rate and propensity to form massive planets in wide orbits, making them ideal targets. The same effects responsible for creating a multitude of detectable planets around massive stars conspire to reduce the number orbiting low-mass stars. However, in the case of a young stellar cluster, where targets are approximately the same age and situated at roughly the same distance, MK-stars can easily dominate the number of detections because of an observational bias related to small number statistics. The degree to which low-mass stars produce the most planet detections in this special case depends upon whether multiple formation mechanisms are at work. Relaxing our assumption that planets with large semimajor axes resemble the RV sample, our calculations suggest that the companions found in ultra-wide orbits around late-type stars in clusters are consistent with a formation channel distinct from that of RV planets. 
\end{abstract}

\section{Introduction}\label{sec:intro}
Precision RV measurements have led to a statistically meaningful sample of extrasolar planet discoveries (http://exoplanets.org) that have shaped much of our understanding of planet formation and evolution. Presently, the RV-detected planets found to orbit most distantly from their host star, 55 Cnc d and HD 190984 b, have semimajor axes of  $\approx6$ AU \citep{fischer_08,santos_10}. As the time baselines of RV surveys grow and more planets are found at incrementally longer periods, a complementary technique, high-contrast imaging, shows promise to supply detections \citep{marois_08,kalas_08,lagrange_10,marois_10} and place tight constraints on the planet population at large separations \citep{nielsen_10} by working from the opposing direction -- outside-in. Of the planets detected directly, the two that orbit closest to their host star, $\beta$ Pictoris b and HR 8799 e, have semimajor axes of $\approx12$ AU and $\approx15$ AU, respectively \citep{lagrange_10,marois_10}. Given the current precisions achieved by RV instruments \citep{marcy_08,mayor_09} and rapid development of coronagraphy and adaptive optics technology \citep{absil_mawet_10,oppenheimer_hinkley_09}, it is clear that this gap in parameter space will continue to narrow and gradually fill. 


While the prospects for continued discoveries are encouraging, it is interesting to consider the origin of planets detected thus far. Both of the aforementioned groups -- planets orbiting close to their host star and those orbiting far away -- have seemingly formed by distinct mechanisms \citep{boley_09}. One currently favored explanation is that core-accretion and other processes related to gravitational disk instabilities are responsible for building each group respectively \citep{pollack_et_al_96,ida_lin_04,boss_97,durisen_07}. This assertion is nominally based on semi-analytic calculations that indicate conditions are unfavorable for planets to form by core-accretion in situ at separations exceeding $\approx 35$ AU, because of a prohibitively slow aggregation of planetesimals \citep{dodson_09,kratter_10}. Taken at face-value, this result implies that either: some of the HR 8799 planets have formed through a different channel than their brethren; the regions over which core-accretion and gravitational instability operate can overlap and the two may possibly conspire; or that the resonant members of a multi-planet system can migrate in unison \citep{crida_09}. The debate regarding how and where Jovian planets typically form is on-going \citep{marois_10,currie_11,ireland_11,boss_11}. These scenarios can be tested by studying the planet population at intermediate and wide separations.

In addition to probing an unexplored parameter space, high-contrast imaging also provides access to photons arriving directly from the planet itself and thus enables characterization studies at a level of detail comparable to transiting planets \citep{seager_10}. Efforts to understand planetary atmospheres are currently underway, but theoretical models are currently in a primitive state, having only a handful of objects with which to study and suffering from model input parameter degeneracies \citep{bowler_10_hr8799b,janson_11,currie_11,madhusudhan_11}. To further inform our theoretical conceptions of planet formation, evolution, and atmospheric physics, it is necessary to obtain a larger sample of companions that reside in orbits beyond $\approx 10$ AU. 

Knowing where to search can help as it may be possible to bias the target selection strategies of planned imaging surveys towards ``planet-enriched" stars. For example, several RV programs have conducted biased surveys by specifically targeting metal-rich stars, leading to a rapid increase in the number of close-in planets \citep{laughlin_00,fischer_05,dasilva_06}. If the number density of RV planets is growing with log orbital period, then a modest extrapolation to separations accessible to high-contrast imaging instruments may provide a reasonable starting point for maximizing the yield of future surveys.

Recent studies of the Doppler RV population show that the planet occurrence rate scales not only with metallicity, but also with stellar mass. \citet{johnson_10_mass_met} have shown that the fraction of stars with planets can be parameterized as a function of stellar mass, $M_*$, and metallicity, $\mbox{[Fe/H]}$, according to:
\begin{equation}
f(M_*,\mbox{[Fe/H]})=k(M_*/M_{\odot})^{\alpha}10^{\beta \mbox{[Fe/H]}},
\label{eqn:mass_met}
\end{equation}
where $k=0.07\pm0.01$, $\alpha=1.0\pm0.3$, and $\beta=1.2\pm0.2$. This result is the first to properly isolate the effects of stellar mass and metallicity on the known planet population and could have important implications for current and future direct-imaging programs, such as the VLT SPHERE \citep{dohlen_06}, Gemini Planet Imager (GPI) \citep{macintosh_GPI_06}, Project 1640 at Palomar \citep{hinkley_11_PASP}, and Subaru SEEDS \citep{tamura_09}. These instruments cover a relatively large range of declinations, and combined, they will observe more than 500 stars in the solar neighborhood over the next several years. It would be useful to identify which kinds of stars are most promising to target.  

Simulations can be used to estimate the degree to which certain spectral-types are likely to host imageable planets. In addition to planet occurrence rates, effects that must be taken into consideration include: the stellar mass function, distribution of local stellar ages and metallicities, planet orbital properties, correlations between stellar properties and planet properties (such as mass and semimajor axis), near-infrared contrast ratio between the star and planet, and, in practice, wavelength range used by the science instrument and AO system wave-front sensor. With exception of the planet semimajor axis distribution, many of these factors are now either known empirically or based on well-constrained observational or theoretical grounds: those related to the number statistics of stars and RV exoplanets in the solar neighborhood are backed by a wealth of high resolution spectroscopic measurements; observation-related trade-offs are appreciated from previous experience with coronagraphic and AO hardware; and, although exoplanet atmospheric models are uncertain at young ages, their systematic over-estimate or under-estimate of the intrinsic brightness of planets in a given band may still be used to calculate direct detection numbers and rates in a relative sense.  

In this paper, we use the results from \citet{johnson_10_mass_met} as input to Monte Carlo simulations to estimate the number of planets that are directly detectable from the ground at near-infrared wavelengths. We consider high-contrast imaging observations that survey target stars using several different representative samples. The results, which are based on the current understanding of planet formation as informed by observations, are tabulated as a function of stellar mass. They may be used to justify and guide target selection for direct imaging programs, in order to maximize their return, and to help understand the theoretical relevance of planet discoveries made in the future at intermediate and large orbital separations. 
 
\section{Description of Monte Carlo Simulations}\label{sec:sims}
To calculate the number of planets detectable by future surveys, we begin by synthesizing a volume-limited sample of stars within 50 pc of the Sun ($\S$\ref{sec:stars}). This distance is large enough to justify the use of statistically significant, empirical results for the properties of stars within the solar neighborhood, yet close enough to yield a reasonable chance for detecting a planet directly with a large-aperture telescope in the near-infrared. It also corresponds to a space-volume wherein all A, F, G stars are bright enough at visible wavelengths to serve as their own natural guide-star. We then populate the stars with planets, according to Eqn.~\ref{eqn:mass_met} and other parameters ($\S$\ref{sec:planets}), and simulate observations with a coronagraph and an ``extreme" AO system ($\S$\ref{sec:instrument}). 


Several different large-scale surveys are considered: a full volume-limited survey where each star within 50 pc is observed ($\S$\ref{sec:volume}); a survey in which stars within this volume are selected by age, to have relatively bright planets, and stellar visual magnitude, for effective AO correction ($\S$\ref{sec:age_bright}); a survey where the selection criterion is modified to incrementally include more distant stars (as in the case of a magnitude-limited sample); and observations of a nearby association that resembles Tucana-Horologium ($\S$\ref{sec:tuc}). We also perform calculations using contrast levels comparable to those achieved by current instruments. 

\subsection{Simulated Stars}\label{sec:stars}
We consider pre-main-sequence, main-sequence, and subgiant stars with mass $0.4 \leq M_* / M_{\odot} \leq 2.6$ as potential targets. Using a number density of 0.11 stars per pc$^3$ \citep{reid_02,latyshev_78}, we simulate a total of 10,081 potential targets (\citet{gray_03}; NStED). This figure assumes a multiplicity factor of 0.33 for the fraction of systems that are binary \citep{raghavan_10} and that $\sim$10$\%$ of those systems have physical separations amenable to high-contrast imaging, including the individual components of wide binaries \citep{holman_wiegert_99}. Each target is assigned a mass, age, and metallicity that conforms to the known bulk statistical properties of stars in the solar neighborhood:\footnote{Ideally, one would acquire stellar properties for actual stars in the solar neighborhood and base a study on their individual properties. However, any attempt to do so would draw upon a heterogeneous collection of measurements that would be incomplete for the volume we wish to consider. To ensure that our estimates of relative detectability as a function of stellar mass are not dominated by systematics or predisposition to certain stellar-types, our study draws upon a synthetic sample.} 

\begin{figure}[!t]
\begin{center}
\includegraphics[height=2.2in]{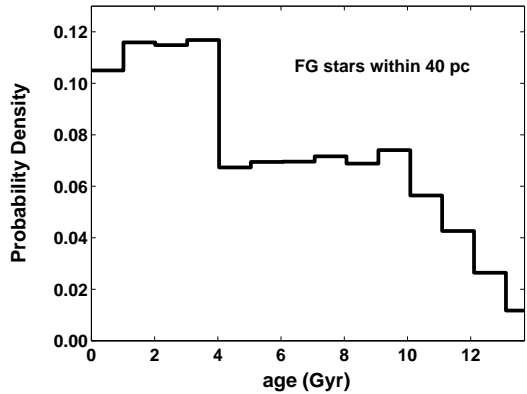}
\caption{Stellar age distribution used for Monte Carlo simulations. Values are drawn for $0.4\leq M_{\odot} \leq 2.6 M_{\odot}$ stars using constraints to ensure consistency with main-sequence and subgiant evolution. Figure adapted from the results of a kinematically unbiased 40 pc volume-limited survey of FG stars by \citet{nordstrom_04}.}\label{fig:age_dist}
\end{center}
\end{figure}

\begin{enumerate} 
\item Masses in the range $0.4 \leq M_* / M_{\odot} \leq 2.6$ (M2-A1 V-IV) are drawn from either a present-day-mass-function (PDMF) for field stars \citep{reid_02}, or an initial mass function (IMF) for stellar clusters \citep{miller_scalo_79}, depending on the simulation. The probability density for masses is given by a broken power-law, $dN_*/dM_* \propto M_*^{\gamma}$. The \citet{reid_02} PDMF has indices $\gamma=-1.35$ for $M_*\leq1.0M_{\odot}$ and $\gamma=-5.2$ for $M_*>1.0M_{\odot}$. The \citet{miller_scalo_79} IMF has indices $\gamma=-1.25$ for $M_*\leq1.0M_{\odot}$, $\gamma=-2.0$ for $1.0 \leq M_*/M_{\odot} \leq 2.0$, and $\gamma=-2.3$ for $M_*>2.0M_{\odot}$. These numbers agree well with NStED database queries.

\item Ages in the range, 4 Myr $< t_{\mbox{\tiny{age}}} < $ 13.7 Gyr, are drawn from a histogram probability distribution that follows a volume-limited sample of well-characterized FG stars within 40 pc based on \citet{nordstrom_04}. Fig.~\ref{fig:age_dist} shows the input used for our code. We assume that the age distribution of all spectral-types follows the same probability distribution. Maximum age constraints are applied to ensure consistency with main-sequence or subgiant evolution for a given star mass. 

\item Metallicities in the range, $-1.0 < \mbox{[Fe/H]} < 0.55$, are drawn from a sliding-Gaussian probability distribution with dispersion, $\sigma_{\small{[Fe/H]}}=0.18$ dex, and center, $\mbox{[Fe/H]}_0$, that changes as a function of age, $t_{\tiny{\mbox{age}}}$, according to $\mbox{[Fe/H]}_0=0.177-0.040\;t_{\tiny{\mbox{age}}}$, in order to simulate the age-metallicity relation derived by \cite{reid_07}, owing to the fact that young stars tend to be metal-rich.  

\item Distances from the Sun, $s$, are drawn from a $dN_* =4\pi s^2 \:ds$ relation. 
\end{enumerate}

{\noindent}We construct $\approx$$10^3$ ensembles (10,081 stars per ensemble) and average the number of planets detected in $0.1M_{\odot}$ wide stellar mass bins to converge to results accurate to $0.1$ planets per bin. 

\subsection{Simulated Planets}\label{sec:planets}
Gas-giant planets are assigned to targets based on the star's mass and metallicity according to Equ.~\ref{eqn:mass_met}. We assume that $28\%$ of host stars have multiple planets, with either 2,3,4 companions each based on random chance \citep{wright_09}. For each planet, we generate orbits using Monte Carlo techniques that take into account projection effects \citep{carson_06,nielsen_08,thalmann_09,beichman_10}: 

\begin{enumerate}
\item Orbits are first constructed in a two dimensional plane from a given semimajor axis and eccentricity. The true anomaly is drawn from a probability distribution that is weighted by the amount of time spent by the planet at a given location in the ellipse. Then, the argument of pericenter and ascending node are chosen randomly and an inclination, $i$, is drawn from a probability distribution, $dn_p/di \propto \mbox{sin}\;i$, where $i=0$ represents face-on orbits. Finally, the orbit is oriented in three dimensions relative to a distant observer using the appropriate rotational transformation matrices \citep{murray_dermott_book}. The angle between the star and planet on the sky is calculated from the projected separation and stellar distance.

\item Semimajor axes in the range $a\leq a_{\mbox{\tiny{max}}}$ are selected considering two different distributions: $dn_p/da \propto a^{-0.61}$ and $dn_p/da \propto a^0$. The first case is representative of a (conventional) planet population that has experienced negligible outward migration or planet-planet scattering, with an index given by the relationship from \citet{cumming_08}, while the second is more applicable to planets that have either formed in-situ at large separation \citep{boley_09,kratter_10} or are dynamically inter-active with their disk or one another \citep{scharf_09,veras_09}. We note that the exact form of the semimajor axis distribution affects the absolute number of planets detected and less-so the relative number. Further, previous authors have found only a weak dependence of their Monte Carlo simulation results (e.g., for the occurrence rate of brown-dwarfs from observations) on the assumed semimajor axis distribution \citep{lafreniere_07,metchev_hillenbrand_09,beichman_10}.

\item The maximal extent of the semimajor axis distribution is scaled as a function of stellar mass, $a_{\mbox{\tiny{max}}}= (M_*/2.5M_{\odot})^{4/9}\tilde{a}$, to account for the theoretically motivated concept that the width of the zone where planets may form increases with stellar mass \citep{kennedy_08}. We later relax this assumption in $\S$\ref{sec:tuc} to isolate its effect on the number of planet detections. Two representative cases are considered: $\tilde{a} = 35, 120$ AU. The first corresponds to a distance beyond which the core-accretion process is expected to become inefficient (see however \citet{currie_11}), and the second corresponds approximately to the location of Fomalhaut b \citep{kalas_08}.

\item The \citet{johnson_10_mass_met} study analyzed planets with semimajor axes $a<2.5$ AU. The constant, $k=0.07$, from Equ.~\ref{eqn:mass_met} therefore represents a lower-limit to the planet occurrence rate. To calculate the overall occurrence rate, $k_e$, we extrapolate the $dn_p/da \propto a^{-0.61}$ semimajor axis distribution to $\tilde{a}=35, 120$ AU (corresponding to 23.3, 79.9 AU for $M_*=1.0M_{\odot}$ stars), finding $k_e=0.17,0.27$ for each case respectively. 

\item We account for the observed paucity of planets within 0.5 AU around A-stars by forbidding semimajor axes in this region for stars with mass $M_*\geq1.5M_{\odot}$ \citep{bowler_10,johnson_10_subgiant}.

\item Eccentricities are drawn from a probability density distribution, $dn_p/de=1-e$. A linear function that falls to zero when $e=1$ well-replicates the number statistics seen in data for exoplanets with orbital periods exceeding 10 days \citep{johnson_09_review}.

\item Planet masses in the range, $0.5 \leq m_p / M_J \leq 15$, are drawn from a power-law distribution, $dn_p/dm_p \propto \; m_p^\gamma$, where $\gamma=-1.4$ \citep{johnson_09_review}. We assume that the index of the true planet mass distribution equals that of the observed $m_p \:\:\mbox{sin}\:i$ distribution. Statistical analyses suggest that this assumption is reasonable for isotropic orbital inclinations in the frame of the planetary system \citep{jorissen_01}, and indirect constraints on the sin$\:i$ distribution, via combined measurements of stellar rotational velocities, periods, and radii, confirm this result for well-aligned planetary systems, i.e., those with small spin-orbit angles \citep{watson_10}. 

\item Finally, we consider cases where A-stars have a propensity to form more massive planets than solar-type stars, by setting $\gamma_A=-1.0$, where $\gamma_A$ is the power-law index as applied to planet masses around $M_*\geq1.5M_{\odot}$ stars \citep{lovis_07,johnson_08}. 
\end{enumerate}
{\noindent}Table~\ref{tab:input} summarizes the number-statistics used for both stellar and planetary properties. 

\begin{table*}[!t]
   \centering   
   \begin{tabular}{lccc}
   \hline
   \hline
      Parameter    &    Description                   & Probability Distribution               &      Reference     \\
      \hline
      \multicolumn{4}{c}{Stars} \\
      \hline
      mass            &    broken power-law         & $\gamma_{\mbox{\tiny{PDMF}}}=-1.35, -5.2   $                               & \cite{reid_02} \\
                          &                                          & $\gamma_{\mbox{\tiny{IMF}}}=-1.25, -2.0, -2.3 $              &  \cite{miller_scalo_79} \\
      age               &   histogram                       &          piece-wise, empirical                          &  \cite{nordstrom_04}   \\
      metallicity     &   sliding Gaussian            & $\sigma_{\tiny{\mbox{[Fe/H]}}}=0.18$ dex   & \cite{reid_07}   \\ 
      \hline
      \multicolumn{4}{c}{Planets} \\
      \hline
      mass             &  power-law   &   $\gamma_{\tiny{\mbox{MKGF}}}=-1.4$, $\;\;\gamma_{\tiny{\mbox{A}}}=-1.4,-1.0$  &  \cite{johnson_09_review}  \\
  semimajor axis &  maximal extent &   $\tilde{a}=35, 120$ AU                                                &  \cite{dodson_09} \\
                            &  power-law         &  $a_{\mbox{\tiny{max}}} = (M_*/2.5M_{\odot})^{4/9}\tilde{a}$  &   \cite{kennedy_08}  \\
                            &  power-law         &  $\gamma=0, -0.61$                                                     &   \cite{cumming_08}  \\
      eccentricity    &  linear                &    $dn_p/de=1-e$                                                           &   \cite{johnson_09_review} \\              
      \hline
   \end{tabular}
   \caption{Input parameters to Monte Carlo simulations. Power-law distribution indices are given by $\gamma$ for the equation $dN/dx \propto x^\gamma$, where $dN/dx$ is the number density as a function of the variable $x$. Stellar PDMF indices correspond to the mass range: $M_*/M_{\odot} \leq1.0$ and $M_*/M_{\odot} > 1.0$ respectively. Stellar IMF indices correspond to the mass range: $M_*/M_{\odot} \leq1.0$, $1.0 < M_*/M_{\odot}\leq 2.0$, and $M_*>2.0M_{\odot}$ respectively. The stellar age distribution is a histogram based on a volume-limited sample of well-characterized stars within 40 pcs. The metallicity distribution is a function of age and has dispersion $\sigma_{\tiny{\mbox{[Fe/H]}}}=0.18$ dex. The maximal extent of planet semimajor axes, $a=\tilde{a}$, is scaled according to stellar mass. The planet eccentricity distribution is applicable to long-period planets discovered by the RV method.}
   \label{tab:input}
\end{table*}

Planet masses are converted to absolute magnitudes in each of the $\lambda=\;$J, H, K bands using the \cite{baraffe_03} ``hot-start" (cond03) and \cite{fortney_08} ``core-accretion" (fort08) evolutionary models. These models make different predictions for the intrinsic brightness of planets at young ages (see \cite{marley_07} for a discussion). We consider ages from 4 Myr to 13.7 Gyrs by interpolating between the values shown in the tables from the above references. To cover this entire range, we also extrapolate each model in time using the slopes for cooling curves from Fig. 1 of \cite{fortney_08}, which are roughly linear in magnitude vs. log($t_{\tiny{\mbox{age}}}$) space, and slopes for cooling curves from \cite{baraffe_03}, which are more appropriate for older planets with low effective temperatures. We also extrapolate the \cite{fortney_08} core-accretion model to $0.5M_J$ and $15M_J$ based on the values provided in the range $1 \leq m_p / M_J \leq 10$. We do not consider the contribution to planet brightness from reflected starlight since thermal emission dominates at near-infrared wavelengths for young bodies with separations exceeding several AU, particularly those that orbit low-mass stars. Stellar masses are converted into absolute magnitudes in each band using \cite{girardi_02} isochrones. The magnitudes of the star and planet are then differenced to calculate contrast ratios. 

While the accuracy of planet evolutionary models is a large uncertainty in the calculations presented, we note that the choice of model is of secondary importance because we are concerned primarily with the \emph{relative} behavior in the number of planets detected as a function of stellar mass. In $\S$\ref{sec:results}, we show that the results from each model are qualitatively similar. We also note that the cond03 and fort08 models are representative of the extrema in brightness predictions at young ages and that the thermal evolution of a planetary atmosphere is likely to fall in between the cooling curves from ``hot-start" initial conditions and those produced from dynamical models of the core-accretion process, based on the physics that they currently incorporate (J. Fortney 2010, private communication). 

\subsection{Simulated Instrument}\label{sec:instrument}
Each target is observed in the J, H, and K bands with a hypothetical $D=8$m telescope that uses a coronagraph. Companions may be detected exterior to the coronagraph inner-working-angle (IWA) which we assume to be 4$\;\lambda/D$. Coronagraphic occulting spots nominally have a radius of $\sim3\;\lambda / D$. In practice, however, the IWA is usually set by the type of speckle-suppression employed \citep{marois_06,biller_10,crepp_10,crepp_11}. We use an effective IWA of 4$\;\lambda/D$ as a representative number. A case where the IWA is set to $3\;\lambda/D$ is considered in $\S$\ref{sec:dist}.

We assume that an ``extreme" AO system generates a dark-hole search region centered on the star \citep{trauger_traub_07,bouchez_09}. The outer-working-angle (OWA), which is governed by the actuator-density of the high-order deformable mirror in the AO system, is set to 26$\;\lambda/D$. PSF subtraction is accounted for by extending the search area by several diffraction widths from the OWA to $1.7\arcsec$ at the detector field-of-view (FOV) edge. It is assumed that the contrast remains constant in this region. The IWA and FOV edge set hard limits to the location where planets may be detected.

Rather than explicitly modeling an optical system, we parameterize the contrast within the dark-hole such that sensitivity improves monotonically with angular separation from the star, falling by 1.5 mags from the IWA to OWA \citep{kataria_10}. We define $C_0$ as the $5\sigma$ contrast at the IWA in the J-band for a $V=5$ star, where $\sigma$ is the speckle noise standard deviation measured relative to the stellar peak intensity. When considering the H and K bands, the instrument IWA and OWA are scaled by $(\lambda/\lambda_0)$ and the contrast within the search region is scaled by $(\lambda_0/\lambda)^2$, where $\lambda_0=\{1.25,1.65,2.20\} \;\mu$m is a reference wavelength at the center of the band, such that longer wavelengths provide wider and deeper dark-holes \citep{malbet_95}. 

Contrast levels are further parameterized based on the target visual magnitude to account for the effects of AO system bandwidth, since fainter stars require longer wave-front sensor integration times which degrades sensitivity. The intensity of scattered starlight is scaled as the square of the rms wave-front error \citep{malbet_95,crepp_09} using calculations from \citet{baranec_08} that include the effects of time delay. The peak planet intensity in each band is likewise modified to account for reductions in Strehl ratio as a function of target brightness. We assume that the AO system has an ability to optimize the wave-front sensor spatial sampling based on the target visual magnitude and assert that targets with $V\leq5$ each achieve the same level of correction. In a similar fashion, we also consider the case of a near-infrared wave-front sensor (NIRWFS) that operates in the J-band, providing access to more low-mass stars \citep{rigaut_92}. We account for the different number of photons per second arriving in the V and J bands for individual targets. 

Planets are detected when their brightness at a given separation exceeds the local $5\sigma$ contrast in either one of the three bands.\footnote{The cond03 and fort08 models are inconsistent with one another in the J and K bands, so we do not perform a bandpass search optimization study (c.f. \citet{agol_07}).} Planet apparent magnitudes must also exceed the noise floor set by the sky background, which we assume to be $m_{\tiny{\mbox{sky}}}=24$ in each band unless otherwise noted, corresponding roughly to the faintest detectable source in $\approx$1 hr of integration time using an 8-10m class telescope with AO \citep{uchimoto_08}. 

When simulating next-generation instruments that use ``extreme" AO, we consider contrast levels: $C_0=5\times10^{-6}$, $5\times10^{-7}$. Most current systems employ low actuator-density deformable mirrors and are limited by non-common-path errors between the AO wave-front sensor and science instrument, limiting contrast levels to $C_0=10^{-3}-10^{-5}$. We also perform runs using $C_0=5\times10^{-5}$ and a wider FOV to compare our calculations to results from current observing programs.   

\section{Results}\label{sec:results}
In the following we present the output of our Monte Carlo simulations for several different high-contrast imaging surveys. Figures generally display two plots: the number of planets detected in each stellar mass bin, $<dn_p / dM_*>$, and the number of planets detected per star in each stellar mass bin, $<(dn_p/dN_*)/dM_*>$, where brackets denote the ensemble average over $\approx10^3$ realizations. For each survey, we consider different values for $\alpha$ and $\beta$ from Eqn.~\ref{eqn:mass_met}, maximal range of semimajor axes, power-law index for planet masses around A-stars, and observations using a NIRWFS. Adjacent curves within each plot nominally have all but one variable held constant to quantify the relative impact of various effects on the number and efficiency of detections. 

\subsection{Volume-Limited Survey}\label{sec:volume}
Volume-limited surveys may be useful when a large number of stars are considered, because they sample a wide range of relevant stellar parameters (mass, age, metallicity), and the results are easier to interpret and less sensitive to uncertainties in the measured properties for specific targets compared to intentionally biased surveys. Simulations of a high-contrast imaging survey of 10,081 A1-M2 V-IV stars within 50 pc (see $\S$\ref{sec:stars} for details) are shown in Figs. 2, 3, using a contrast of $C_0=5\times10^{-7}$ with the cond03 and fort08 planet evolutionary models respectively. 

\subsubsection{Planet Detections}
We find that the number of directly detectable gas-giant planets in the solar neighborhood follows a distribution that is multi-modal when plotted against stellar mass, with peaks located at $M_*\lesssim0.4M_{\odot}$, $M_* \approx 1.0M_{\odot}$, and $M_* \approx 1.6M_{\odot}$. The middle peak is common to all cases considered, while the last peak occurs only when taking into consideration correlations between star mass and planet mass. The first peak in the number of planet detections is seen in cases where the planet occurrence rate scales with star mass (Equ. 1), while its existence is only implicit in other situations (see Fig. 2). 

These results are driven by the tradeoffs listed in $\S$\ref{sec:intro}. For late-type stars, the number of detections is governed by the competing effects of an increasing stellar mass function, favorable star-to-planet brightness ratios, a diminished planet occurrence rate, and poor AO correction. To provide a fiducial measure for the number of available targets, we have over-plotted the \citet{reid_02} PDMF, scaled down by a large multiplicative factor to fit inside of Figs. 2, 3. The number of detections closely follows this shape for early M-type and late K-type stars. In this regime, targets are sufficiently faint such that nearly all young Jovian planets in wide orbits have an excellent chance of being imaged directly, provided they are located inside of the instrument FOV. The factor that limits the number of detections is that few nearby late-type stars are also young.\footnote{See \citet{liu_09} for a discussion of the ``missing young M-dwarfs problem".}

The number of planet detections begins to turn over (decrease) for M-stars when the planet occurrence rate obeys Equ. 1, owing to the fact that low-mass stars have few Jovian planets. Further, insufficient compensation for atmospheric turbulence for the faintest stars reduces the Strehl ratio and can lead to prohibitively small companion peak intensities when observed against the NIR sky background. As a result of these effects, the number of detectable planets must approach zero at $M_* < 0.4M_{\odot}$, even when using a NIRWFS. However, depending on the formation scenario, for instance whether or not $\alpha$ and $\beta$ have non-zero values, a peak may not occur until much lower masses, possibly in the brown dwarf regime. 

Solar-type stars are sufficiently bright for ``extreme" AO correction, but the number of available targets decreases sharply for $M_*>1.0M_{\odot}$ where the PDMF in the solar neighborhood experiences a break in the power law index. A local maximum in the number of planet detections occurs at $M_*\approx1.0M_{\odot}$ independent of simulation input. The number of planet detections falls monotonically for $M_*>1.0M_{\odot}$, except when there exist correlations between star mass and planet mass. We find that changing the index of the planet mass power law from $\gamma_A=-1.4$ to $\gamma_A=-1.0$ for $M_* \geq 1.5M_{\odot}$ generates another peak near $M_*\approx1.6M_{\odot}$ by amplifying the number of detections around massive stars. 

\begin{figure*}[!ht]
\begin{center}
\includegraphics[height=2.6in]{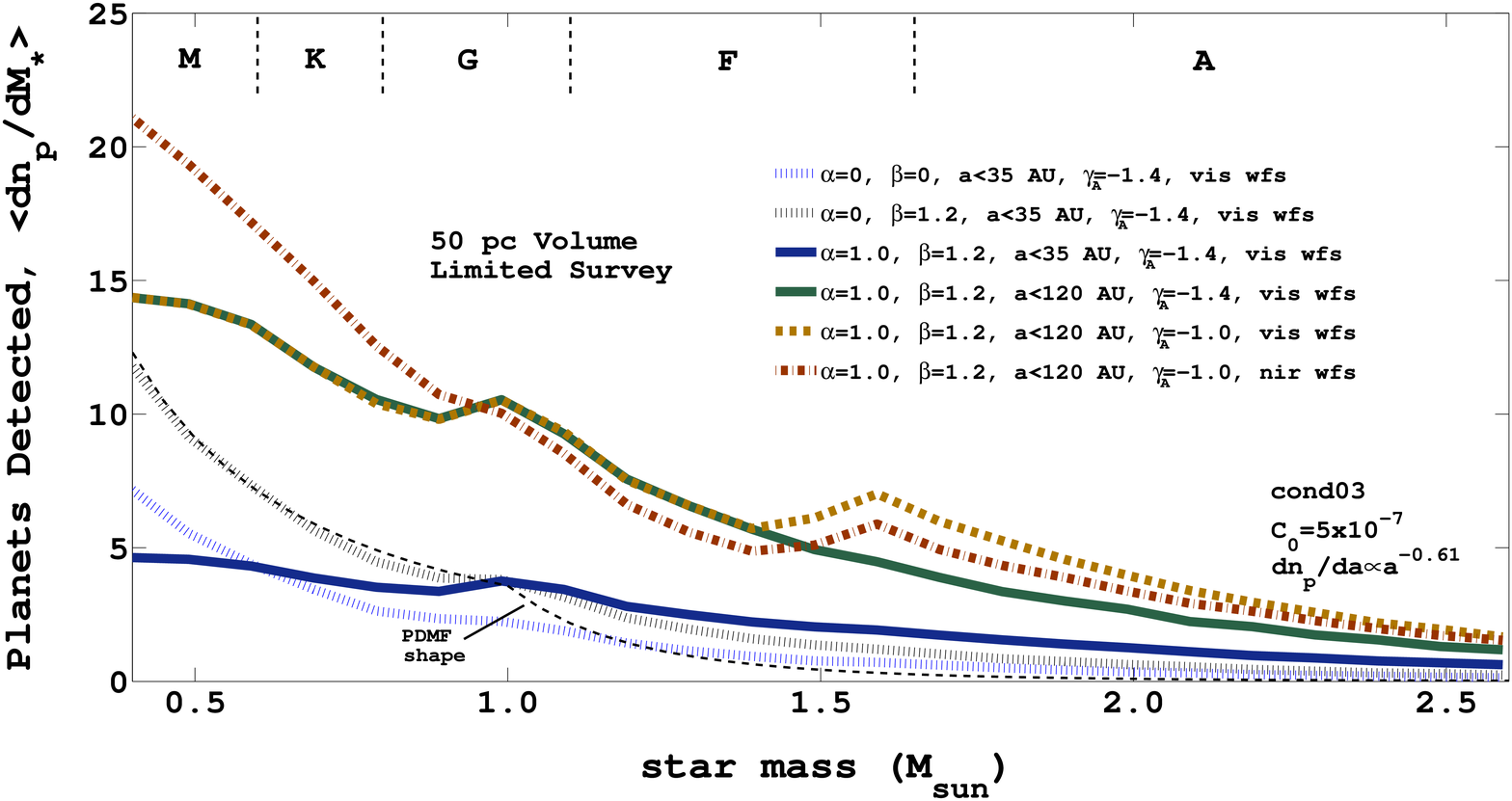}
\includegraphics[height=2.6in]{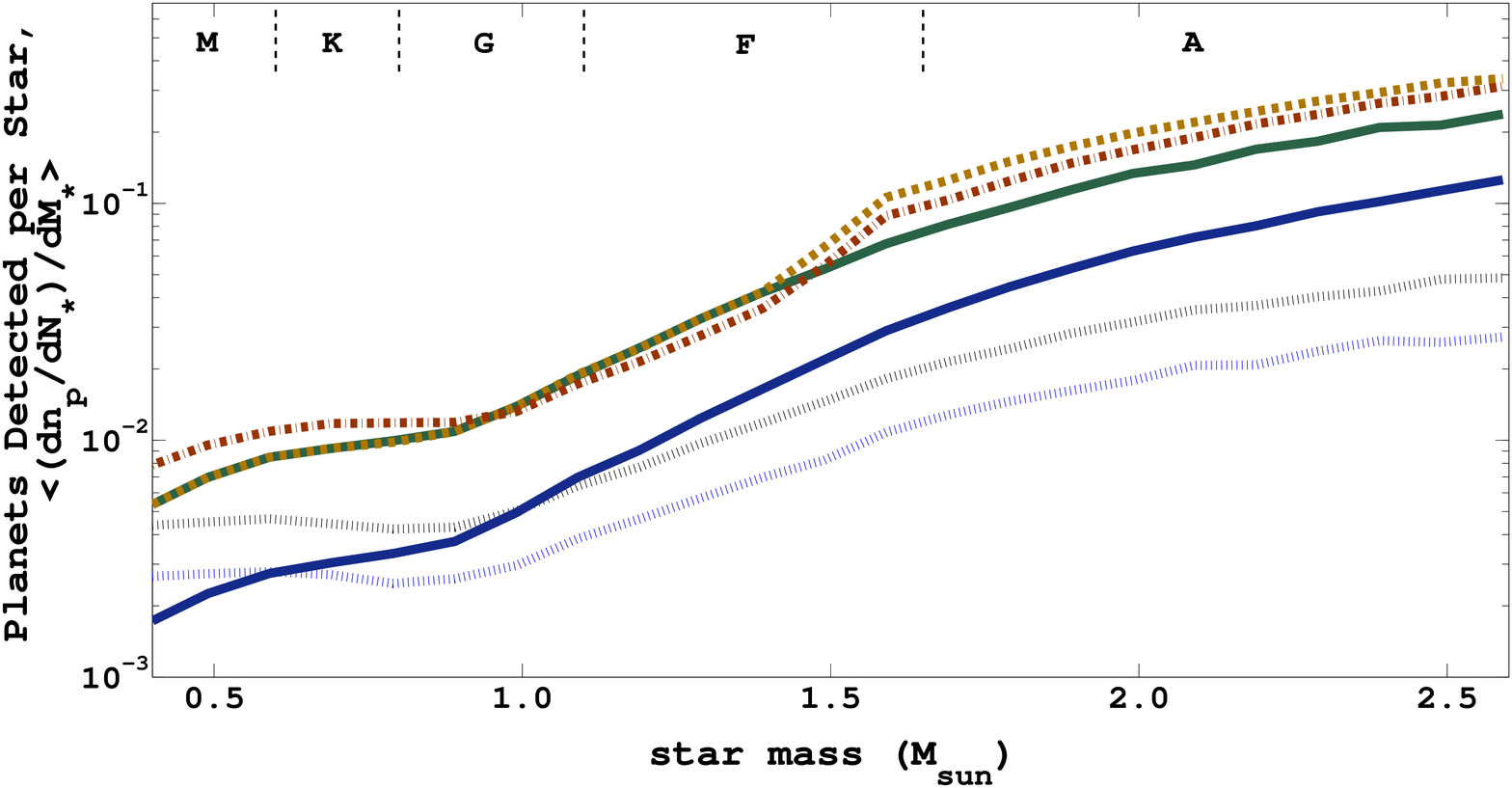}
\caption{Monte Carlo simulation results of a 50 pc ground-based volume-limited survey of 10,081 stars using the planet occurrence-rates from \citet{johnson_10_mass_met}, stellar and planetary distributions in Table~\ref{tab:input}, an 8m telescope with ``extreme" AO and a NIR coronagraph that generate contrast levels of $C_0=5\times10^{-7}$, and the cond03 planet thermal evolutionary models. The overall number of detectable planets (top) is governed primarily by the stellar mass function (PDMF) for late spectral types, while observing efficiency, or the number of planet detections per star (bottom), lies heavily in favor of early spectral types as a consequence of their intrinsic youth and visual brightness for AO correction. Including correlations between star mass and planet  occurrence rate and other properties exacerbates the difference. Legends are the same for both plots. Main-sequence spectral-types are labeled on the top horizontal axis for reference. Each curve is sampled by 21 data points (histogram bins) in stellar mass.}
\end{center}\label{fig:vol_bar}
\end{figure*} 

\begin{figure*}[!ht]
\begin{center}
\vspace{0.3in}
\includegraphics[height=2.6in]{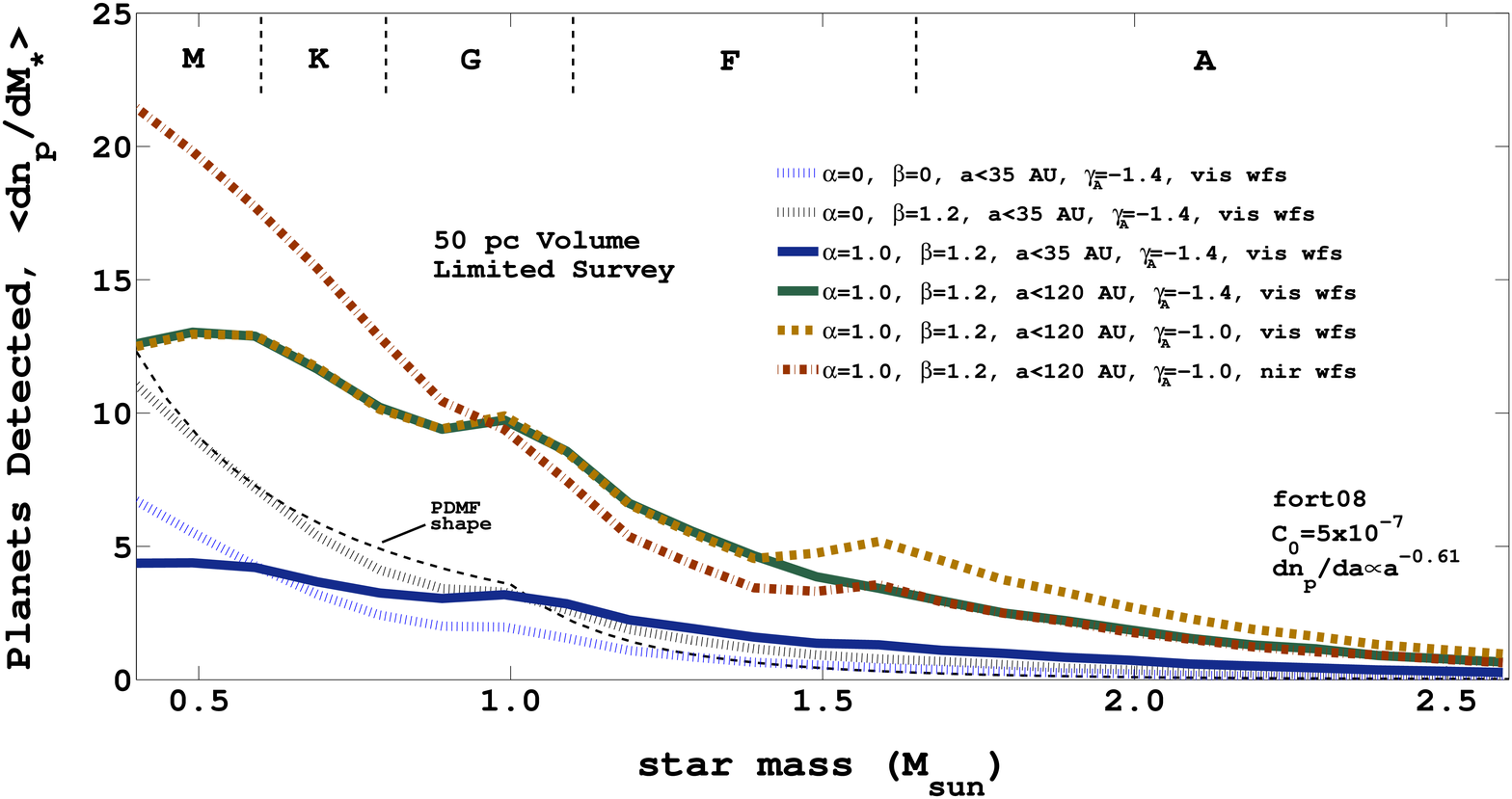}
\includegraphics[height=2.6in]{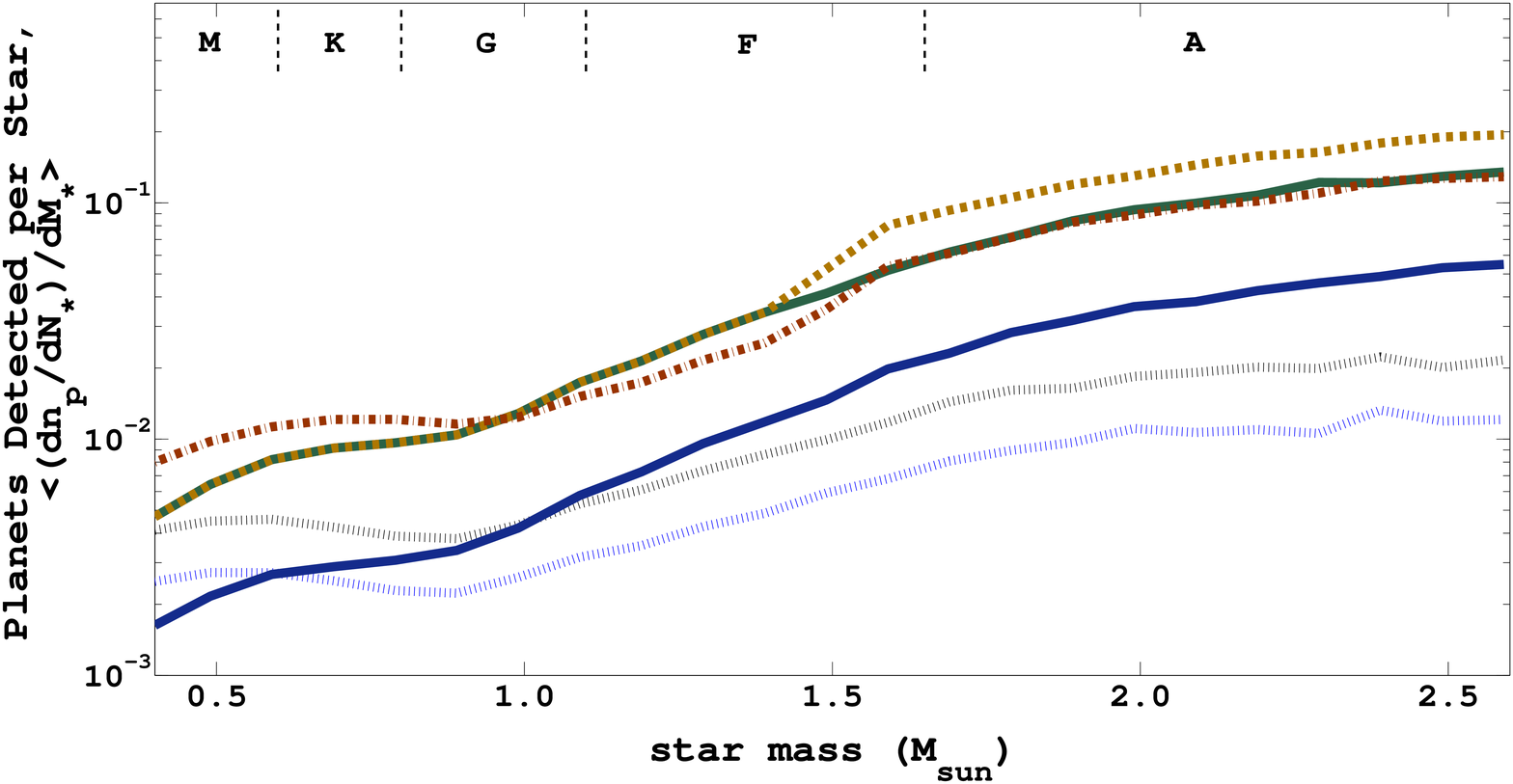}
\caption{Same as Fig. 2 using the fort08 evolutionary models. Fewer planets are detected compared to the ``hot-start" models but the relative number and rate of detections are qualitatively similar.}
\end{center}\label{fig:vol_fort}
\end{figure*}

\begin{table*}[!t]
  \centering 
   \begin{tabular}{cccccc|c|c|c|c}
   \hline
    \multicolumn{10}{c}{Volume-Limited Survey} \\
   \hline
   \multicolumn{10}{c}{$dn_p/da \propto a^{-0.61}$, $k_e=\{0.17,0.27\}$} \\
   \hline 
\multicolumn{6}{c|}{input parameters}  &  \multicolumn{4}{c}{stellar mass range ($M_{\odot}$)}  \\
\hline
$\alpha$ & $\beta$ & $C_0$ & $\gamma_A$ & $\tilde{a}$ & WFS & 0.40-0.95 & 0.95-1.50 & 1.50-2.05 &  2.05-2.60  \\
 \hline
 0  &  0  &  $5\times10^{-6}$  &   -1.4   & 35   &  vis  & 6.0 $\:$ 1.4  & 2.2 $\:$ 0.2 & 0.9 $\:$ 0.0 & 0.4 $\:$ 0.0  \\
0  &  1.2 &  $5\times10^{-6}$ &  -1.4   & 35   &  vis  & 10  $\:$ 2.4  & 3.7 $\:$ 0.4 & 1.6 $\:$ 0.0 & 0.7 $\:$ 0.0 \\
1.0 & 1.2 & $5\times10^{-6}$ &  -1.4    & 35  &  vis &  6.2 $\:$ 1.3 & 4.6 $\:$ 0.4 & 2.8 $\:$ 0.1 & 1.5 $\:$ 0.0 \\
1.0 & 1.2 & $5\times10^{-6}$ &  -1.4   & 120 &  vis  & 20 $\:$ 7.2 & 12 $\:$ 2.0 & 6.8 $\:$ 0.5 & 3.3 $\:$ 0.1 \\
\bf{1.0} & \bf{1.2} & \boldmath{$5\times10^{-6}$} &  \bf{-1.0}   & \bf{120} &  \bf{vis}  & 20 $\:$ 7.2 & 12 $\:$ 2.0 & 11 $\:$ 0.6 & 5.4 $\:$ 0.1 \\
1.0 & 1.2 & $5\times10^{-6}$ &  -1.0   & 120 &  nir  & 29 $\:$ 18 & 11 $\:$ 1.3 & 8.6 $\:$ 0.2 & 4.4 $\:$ 0.0 \\
\hline
0  &  0  &  $5\times10^{-7}$  &   -1.4   & 35   &  vis  & 21 $\:$ 20 & 7.9 $\:$ 6.3 & 3.0 $\:$ 1.9 & 1.2 $\:$ 0.6 \\
0  &  1.2 &  $5\times10^{-7}$ &  -1.4   & 35   &  vis  & 35 $\:$ 34 & 13 $\:$ 11 & 5.1 $\:$ 3.2 & 2.1 $\:$ 1.1 \\
1.0 & 1.2 & $5\times10^{-7}$ & -1.4   & 35    &  vis  & 22 $\:$ 21 & 16 $\:$ 12 & 8.8 $\:$ 5.6 & 4.7 $\:$ 2.3 \\
1.0 & 1.2 & $5\times10^{-7}$ &  -1.4   & 120 &  vis  & 66 $\:$ 63 & 42 $\:$ 37 & 20 $\:$ 15 & 9.4 $\:$ 6.0 \\
\bf{1.0} & \bf{1.2} & \boldmath{$5\times10^{-7}$} &  \bf{-1.0}   & \bf{120} &  \bf{vis}  & 66 $\:$ 63 & 42 $\:$ 37 & 30 $\:$ 22 & 14 $\:$ 8.6 \\
1.0 & 1.2 & $5\times10^{-7}$ &  -1.0   & 120 &  nir  & 84 $\:$ 86 & 37 $\:$ 31 & 25 $\:$ 15 & 12 $\:$ 5.7 \\
\hline  
\hline
\multicolumn{10}{c}{$dn_p/da \propto a^0$, $k_e=\{0.17,0.27\}$} \\
\hline 
\multicolumn{6}{c|}{input parameters}  &  \multicolumn{4}{c}{stellar mass range ($M_{\odot}$)} \\
\hline
$\alpha$ & $\beta$ & $C_0$ & $\gamma_A$ & $\tilde{a}$ & WFS & 0.40-0.95 & 0.95-1.50 & 1.50-2.05 &  2.05-2.60  \\
 \hline
0  &  0  &  $5\times10^{-6}$  &   -1.4   & 35   &  vis  & 9.5  $\:$ 2.3 & 3.5 $\:$ 0.3 & 1.3 $\:$ 0.0 & 0.6 $\:$ 0.0  \\
0  &  1.2 &  $5\times10^{-6}$ &  -1.4   & 35   &  vis  & 16  $\:$ 4.0 & 6.0 $\:$ 0.6 & 2.3 $\:$ 0.1 & 1.0 $\:$ 0.0 \\
1.0 & 1.2 & $5\times10^{-6}$ &  -1.4    & 35  &  vis & 9.9 $\:$ 2.2 & 7.2 $\:$ 0.8 & 4.0 $\:$ 0.1 & 2.5 $\:$ 0.1 \\
1.0 & 1.2 & $5\times10^{-6}$ &  -1.4   & 120 &  vis  & 28 $\:$ 11 & 17 $\:$ 3.2 & 8.0 $\:$ 0.9 & 4.2 $\:$ 0.2 \\
\bf{1.0} & \bf{1.2} & \boldmath{$5\times10^{-6}$} &  \bf{-1.0}   & \bf{120} &  \bf{vis}  & 28 $\:$ 11 & 17 $\:$ 3.2 & 14 $\:$ 1.0 & 6.5 $\:$ 0.2 \\
1.0 & 1.2 & $5\times10^{-6}$ &  -1.0   & 120 &  nir  & 42 $\:$ 28 & 15 $\:$ 2.1 & 10 $\:$ 0.2 & 5.7 $\:$ 0.0 \\
\hline
0  &  0  &  $5\times10^{-7}$  &   -1.4   & 35   &  vis  & 34 $\:$ 32 & 13 $\:$ 10 & 4.2 $\:$ 2.8 & 1.6 $\:$ 0.9 \\
0  &  1.2 &  $5\times10^{-7}$ &  -1.4   & 35   &  vis  & 57 $\:$ 54 & 21 $\:$ 17 & 7.3 $\:$ 4.8 & 2.9 $\:$ 1.6 \\
1.0 & 1.2 & $5\times10^{-7}$ & -1.4   & 35    &  vis &  35 $\:$ 33 & 25 $\:$ 20 & 13 $\:$ 8.3 & 6.6 $\:$ 3.6 \\
1.0 & 1.2 & $5\times10^{-7}$ &  -1.4   & 120 &  vis  & 94 $\:$ 90 & 57 $\:$ 51 & 24 $\:$ 19 & 11 $\:$ 7.6 \\
\bf{1.0} & \bf{1.2} & \boldmath{$5\times10^{-7}$} &  \bf{-1.0}  & \bf{120} &  \bf{vis} & 94 $\:$ 90 & 57 $\:$ 51 & 36 $\:$ 28 & 16 $\:$ 11 \\
1.0 & 1.2 & $5\times10^{-7}$ &  -1.0   & 120 &  nir  & 117 $\:$ 120 & 51 $\:$ 43 & 31 $\:$ 20 & 14 $\:$ 7.4 \\
\hline
\hline
\multicolumn{6}{c|}{Number of Stars Observed} & 7,913 & 1,885 & 230 & 53 \\
\hline
\end{tabular}
\caption{Number of planet detections (ensemble average) for a 50 pc ground-based volume-limited survey of 10,081 stars over several stellar mass bins. Entries for cond03 and fort08 models are shown in the left and right of each column respectively. Two different semimajor axis distributions are considered. The gas-giant planet occurrence rate, $k_e=\{0.17,0.28\}$, is applied to semimajor axis distributions truncated at $\tilde{a}=35$ AU and $\tilde{a}=120$ AU respectively. Calculations allow for multi-planet systems. Rows having parameters that may be considered as the baseline case for an extrapolation of the RV planet population are bold-faced.}
   \label{tab:npl_vol}
\end{table*}


Increasing the range of semimajor axes from $\tilde{a}=35$ AU to $\tilde{a}=120$ AU enhances the number of planet detections for most stellar masses. However, the most massive stars ($M_*\approx2.5M_{\odot}$) may experience the smallest relative gain since we have scaled the maximal extent of semimajor axes by star mass ($\S$\ref{sec:sims}). The instrument FOV edge is $1.7\arcsec$ from the star and planets with $a=120$ AU and near face-on orbits around A-stars within 50 pc will subtend a wide angle on the sky and lie outside of the search region. This effect, though minor, will also impact next generation high-contrast imaging instruments when observing stars with close proximity to the Sun \citep{crepp_09}. For example, the GPI has a FOV width of 2.8"$\times\:$2.8" and will therefore not be able to detect and characterize Fomalhaut b in default operating mode, i.e., with the star occulted by the coronagraph. 

We find that a NIRWFS operating in the J-band can increase the number of planet detections around low-mass stars by tens of percent, providing access to more targets with red spectral energy distributions. Presumably, a laser-guide-star system can serve a similar role \citep{wizinowich_10}. Stars with blue spectral energy distributions experience only a modest loss. We note that such considerations are less important for space observations because the stellar wavefront may be sensed in the image plane by the science camera at the same wavelengths \citep{giveon_07}. 

The relative number of detections as a function of stellar mass is qualitatively similar between the top panels of Fig. 2 and Fig. 3, but the absolute numbers differ because the cond03 and fort08 models make different estimates for the brightness of young planets. While the models generally agree on timescales of several Gyrs, the brightness difference can be an order of magnitude for massive planets with $t_{\mbox{\tiny{age}}}\lesssim100$ Myr and two orders of magnitude at $t_{\mbox{\tiny{age}}}\lesssim10$ Myr \citep{marley_07}. The disparity between model predictions is evident for bright stars, where contrast matters most, and with instruments that generate contrast levels of $C_0\geq5\times10^{-6}$. In this regime, where only the youngest and most massive planets are detectable, the cooling tracks are most discrepant, requiring, for example, approximately 1 Gyr to converge for an $m_p=8M_J$ planet \citep{fortney_08}. Table \ref{tab:npl_vol} shows the total number of planet detections for each case considered in different stellar mass bins, using $C_0=5\times10^{-6}$ and $C_0=5\times10^{-7}$ and two different semimajor axis distributions. 

\subsubsection{Planet Detection Rates}
Importantly, the shape of the planet detection curves do not follow the PDMF at high star masses, but instead rise above the (normalized) stellar mass function for all cases considered, indicating significantly higher detection efficiencies compared to low-mass stars. Dividing the number of detections in each stellar mass bin by the number of stars observed in each stellar mass bin to calculate a detection rate, $<(dn_p/dN_*)/dM_*>$, we find that the likelihood for imaging a planet favors massive stars by more than an order of magnitude over low-mass stars. Detection rate curves are shown on a logarithmic scale in the bottom panels of Fig. 2, 3.

Massive stars have a high planet occurrence-rate and are comparatively young while on the main-sequence and early subgiant branch. Given their youth, they have bright planets and are also slightly more metal-rich than low-mass stars (on average), which leads to an additional increase (coupling) in the planet occurrence-rate through Eqn.~\ref{eqn:mass_met}. Massive stars are also expected to have wide planet formation zones and are sufficiently bright at visible wavelengths to serve as their own guide star. As stars with progressively higher masses are observed, these effects combine in a non-linear way to increase the efficiency of direct imaging observations. As a result, A-stars to yield a comparable number of planet detections as M-stars despite comprising a much smaller fraction of potential targets in the sky.  

While volume limited surveys may at first seem appealing because they involve a systematic search of the Sun's closest neighbors, they are inefficient unless targets are restricted to high stellar mass. For instance, using the cond03 models with $C_0=5\times10^{-7}$, $\tilde{a}=120$ AU, $\gamma_A=-1.0$, $\alpha=1.0$, and $\beta=1.2$, the average number of $M_*=1.5M_{\odot}$ stars within 50 pc that need to be observed to discover one Jovian planet in an unbiased survey is $\approx$15, while the corresponding number at $M_*=0.7M_{\odot}$ is $\approx$85 with a NIRWFS, and $\approx$107 without. In $\S$\ref{sec:age_bright} and $\S$\ref{sec:tuc}, we consider surveys where stars are pre-selected for youth to make a more interesting comparison and determine whether this trend with stellar mass is upheld.

\subsection{Age and Brightness Selected Survey}\label{sec:age_bright}
Most current and planned surveys select targets to maximize the chances for discovery. In this section, we consider another large-scale high-contrast imaging program in which stars from the 50 pc volume-limited sample are chosen to have apparent magnitudes, $V<9$ ($J<8$ with a NIRWFS), and ages, $t_{\mbox{\tiny{age}}}<1$ Gyr. An age cut of 1 Gyr is convenient because it corresponds to a value in which discrimination between marginally younger or older stars may be done with reasonable consistency based on their level of chromospheric activity \citep{noyes_84}. Further, low-mass planets that are 1 Gyr old have likely cooled to a point beyond which direct detection becomes extremely challenging. Selecting bright and young stars from the volume-limited sample yields a total of 610 targets when using a VISWFS and 760 targets when using a NIRWFS. Our results for simulated observations of each star with $C_0=5\times10^{-7}$ using the same techniques and parameters described in $\S$\ref{sec:sims} and $\S$\ref{sec:volume} are shown in Figs. 4, 5. Table 3 displays the number of planet detections for various contrast levels and two different semimajor axis distributions. 

\subsubsection{Planet Detections}
We find that the number of planet detections rises steadily with increasing stellar mass from M-dwarfs to G-dwarfs and, like a volume-limited survey, exhibits a peak near $1.0M_{\odot}$. However, unlike a volume-limited survey, few detections are made at low stellar masses on account of a dearth of available targets. This result is a consequence of the local age distribution within the solar neighborhood (Fig. 1). Only $\lesssim15\%$ of nearby FGK stars are younger than 1 Gyr, and M-stars are older yet \citep{mamajek_hillenbrand_08}. 

\begin{figure*}[!t]
\begin{center}
\includegraphics[height=2.6in]{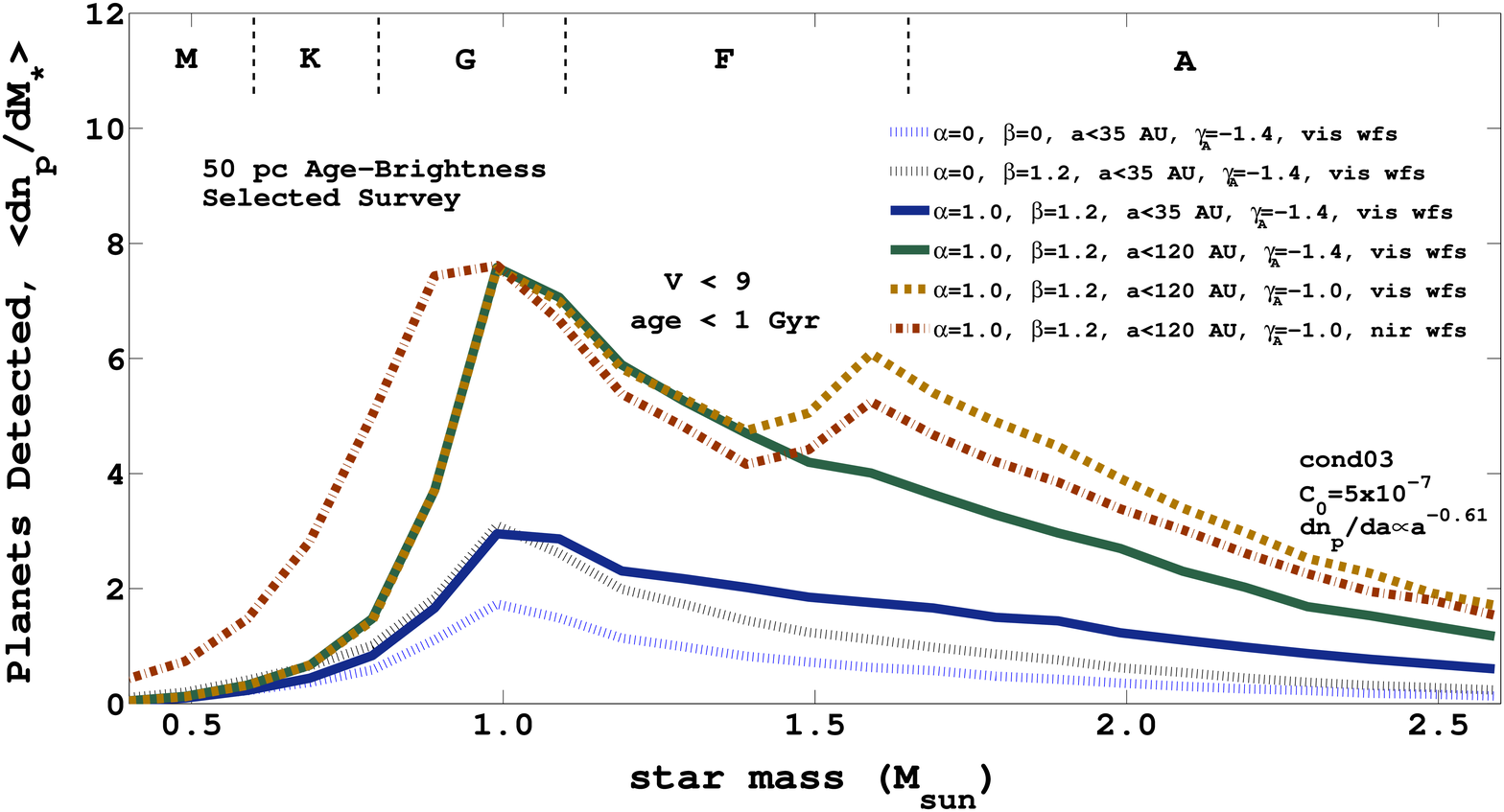}
\includegraphics[height=2.6in]{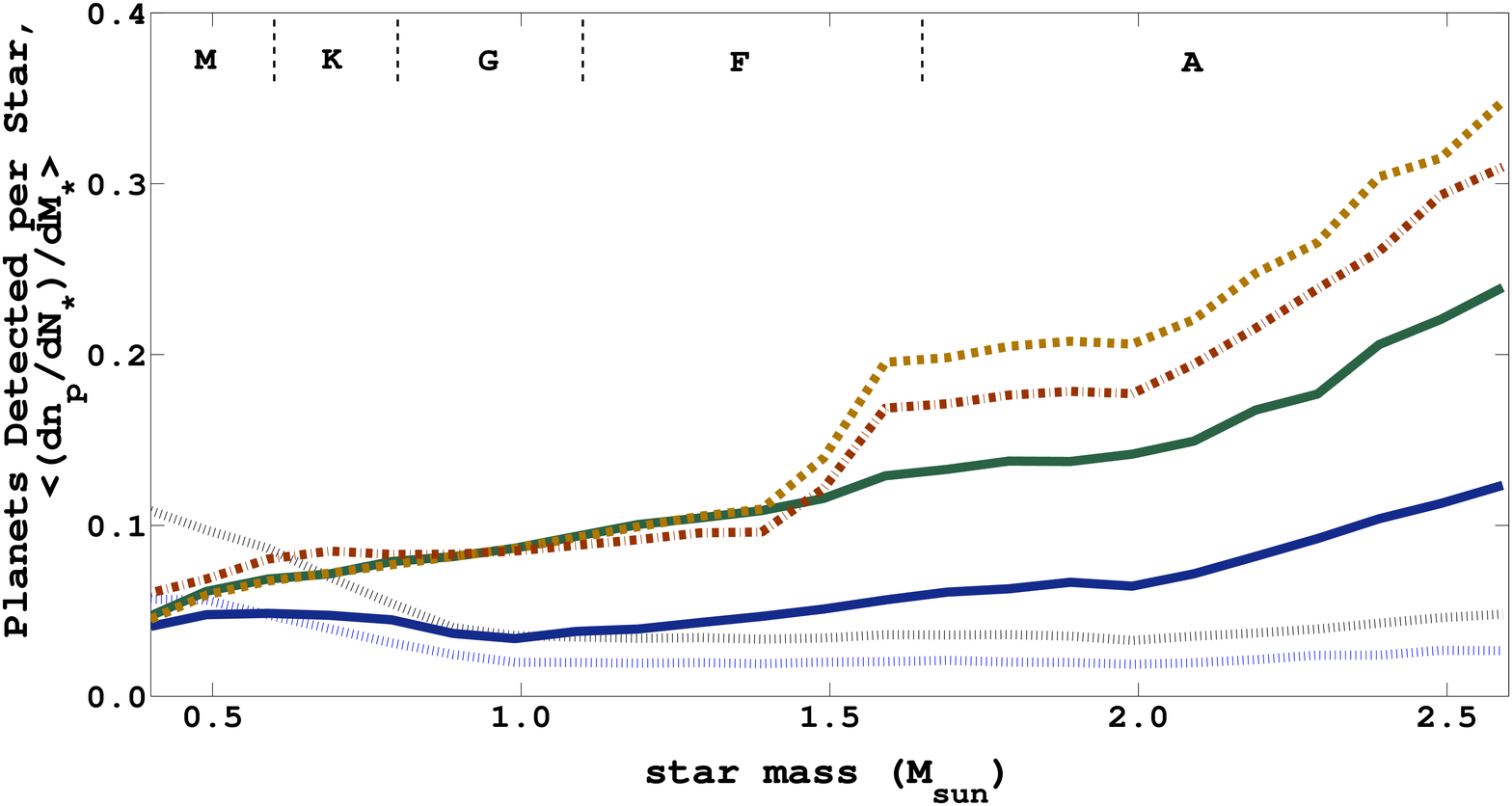}
\caption{Monte Carlo simulation results of a 50 pc survey of 610 stars (760 with NIRWFS) with age, $t_{\tiny{\mbox{age}}}< 1$ Gyr and $V<9$ ($J<8$) using the planet occurrence rates from \citet{johnson_10_mass_met}, stellar and planetary distributions from Table 1, an 8m telescope with ``extreme" AO and a near-infrared coronagraph that generate contrast levels of $C_0=5\times10^{-7}$, and the \citet{baraffe_03} ``hot-start" planet evolutionary models. A-stars yield a comparable number of detections to FGKM stars and have elevated detection rates even when stars are pre-selected for youth and brightness. Legends are the same for both plots. Main-sequence spectral-types are labeled on the top horizontal axis for reference. Each curve is sampled by 21 data points in stellar mass. Color-scheme is the same as the volume-limited case.}
\end{center}\label{fig:age_bright_bar}
\end{figure*}

\begin{figure*}[!t]
\begin{center}
\includegraphics[height=2.6in]{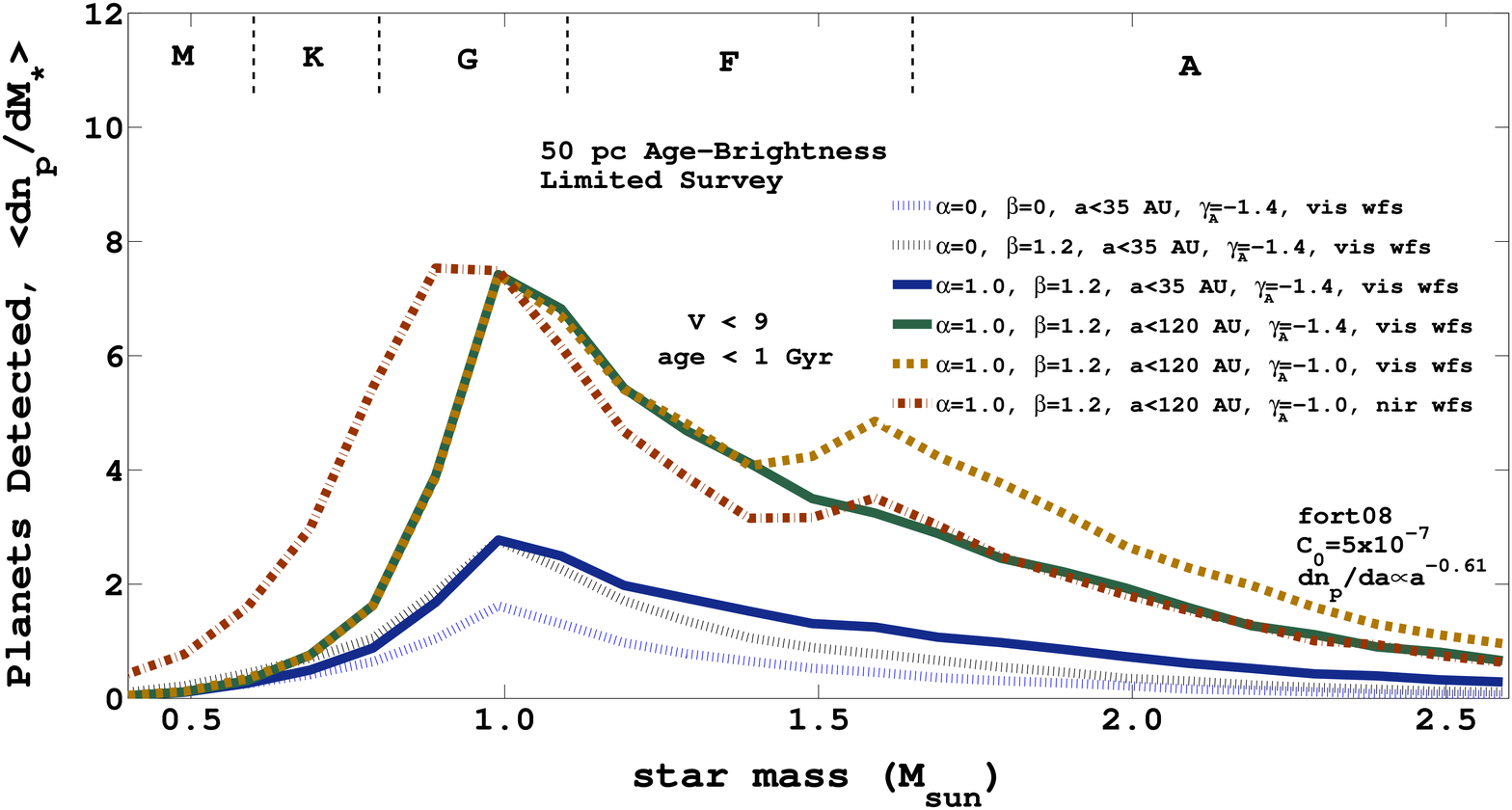}
\includegraphics[height=2.6in]{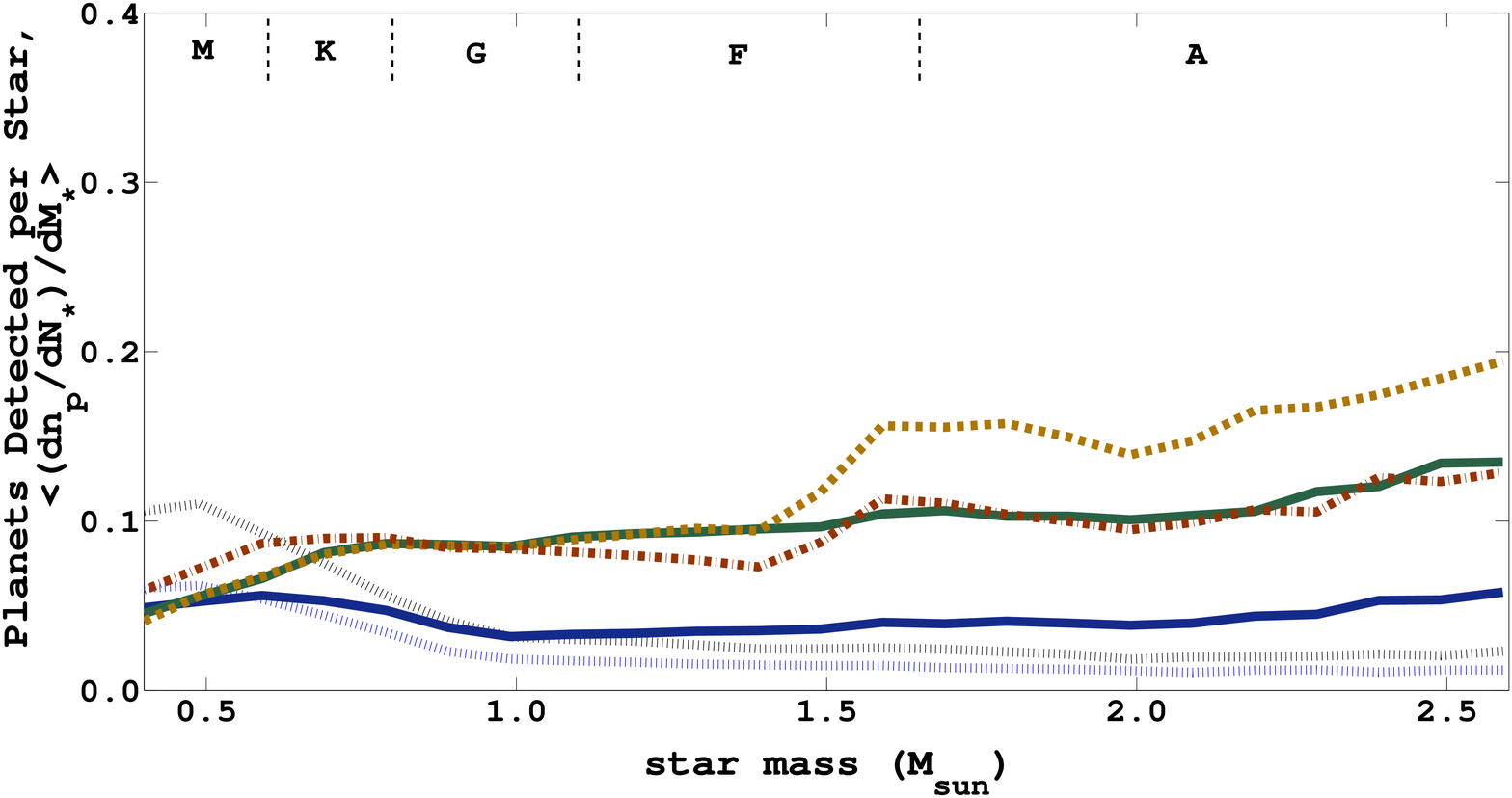}
\caption{Same as Fig. 4 using the Fortney et al. 2008 ``core-accretion" models.}
\end{center}\label{fig:age_bright_fort}
\end{figure*}

\begin{table*}[!t]
\vspace{-0.7in}
  \centering 
   \begin{tabular}{cccccc|c|c|c|c}
   \hline    
   \multicolumn{10}{c}{Age and Brightness Selected Survey} \\
   \hline
   \multicolumn{10}{c}{$dn_p/da \propto a^{-0.61}$, $k_e=\{0.17,0.27\}$} \\
   \hline 
   \multicolumn{6}{c|}{input parameters}  &  \multicolumn{4}{c}{stellar mass range ($M_{\odot}$)} \\
   \hline
$\alpha$ & $\beta$ & $C_0$ & $\gamma_A$ & $\tilde{a}$ & WFS & 0.40-0.95 & 0.95-1.50 & 1.50-2.05 &  2.05-2.60 \\
\hline
0  &  0  &  $5\times10^{-5}$  &   -1.4   & 35   &  vis  &  0.4 $\:$ 0.1 & 0.8 $\:$ 0.0 & 0.3 $\:$ 0.0 & 0.1 $\:$ 0.0  \\
0  &  1.2 &  $5\times10^{-5}$ &  -1.4   & 35   &  vis  &  0.7 $\:$ 0.1 & 1.2 $\:$ 0.0 & 0.6 $\:$ 0.0 & 0.2 $\:$ 0.0 \\
1.0 & 1.2 & $5\times10^{-5}$ &  -1.4    & 35  &  vis &  0.5 $\:$ 0.1 & 1.5 $\:$ 0.0 & 1.0 $\:$ 0.0 & 0.5 $\:$ 0.0 \\
1.0 & 1.2 & $5\times10^{-5}$ &  -1.4   & 120 &  vis  & 1.7 $\:$ 0.4 & 5.8 $\:$ 0.2 & 3.5 $\:$ 0.0 & 2.1 $\:$ 0.0 \\
\bf{1.0} & \bf{1.2} & \boldmath{$5\times10^{-5}$} &  \bf{-1.0}   & \bf{120} &  \bf{vis}  & 1.7 $\:$ 0.4 & 5.8 $\:$ 0.2 & 5.9 $\:$ 0.0 & 3.2 $\:$ 0.0 \\
1.0 & 1.2 & $5\times10^{-5}$ &  -1.0   & 120 &  nir  & 4.6 $\:$ 1.3 & 5.3 $\:$ 0.1  & 5.0 $\:$ 0.0 & 2.8 $\:$ 0.0 \\
 \hline
0  &  0  &  $5\times10^{-6}$  &   -1.4   & 35   &  vis  & 1.1 $\:$ 0.5  & 2.2 $\:$ 0.2 & 0.9 $\:$ 0.0 & 0.4 $\:$ 0.0   \\
0  &  1.2 &  $5\times10^{-6}$ &  -1.4   & 35   &  vis  &  2.0 $\:$ 0.9 & 3.7 $\:$ 0.4 & 1.6 $\:$ 0.0 & 0.7 $\:$ 0.0 \\
1.0 & 1.2 & $5\times10^{-6}$ &  -1.4    & 35  &  vis &  1.5 $\:$ 0.6 & 4.4 $\:$ 0.5 & 2.7 $\:$ 0.1 & 1.5 $\:$ 0.1 \\
1.0 & 1.2 & $5\times10^{-6}$ &  -1.4   & 120 &  vis  & 2.9 $\:$ 1.4 & 12 $\:$ 2.0 & 6.7 $\:$ 0.5 & 3.3 $\:$ 0.1 \\
\bf{1.0} & \bf{1.2} & \boldmath{$5\times10^{-6}$} &  \bf{-1.0}   & \bf{120} &  \bf{vis}  & 2.9 $\:$ 1.4 & 12 $\:$ 2.0 & 10 $\:$ 0.6 & 5.4 $\:$ 0.1 \\
1.0 & 1.2 & $5\times10^{-6}$ &  -1.0   & 120 &  nir  & 8.4 $\:$ 4.5 & 11 $\:$ 1.3  & 8.6 $\:$ 0.2  & 4.4 $\:$ 0.0 \\
\hline
0  &  0  &  $5\times10^{-7}$  &   -1.4   & 35   &  vis  & 2.6 $\:$ 2.7 & 6.5 $\:$ 5.5 & 2.8 $\:$ 1.9 & 1.2 $\:$ 0.6 \\
0  &  1.2 &  $5\times10^{-7}$ &  -1.4   & 35   &  vis  & 4.4 $\:$ 4.6 & 11 $\:$ 9.5 & 4.9 $\:$ 3.2 & 2.1 $\:$ 1.1 \\
1.0 & 1.2 & $5\times10^{-7}$ & -1.4   & 35    &  vis &  3.6 $\:$ 3.7 & 13 $\:$ 11 & 8.5 $\:$ 5.5 & 4.7 $\:$ 2.3 \\
1.0 & 1.2 & $5\times10^{-7}$ &  -1.4   & 120 &  vis  & 6.9 $\:$ 7.3 & 32 $\:$ 30 & 19 $\:$ 14 & 9.4 $\:$ 5.9 \\
\bf{1.0} & \bf{1.2} & \boldmath{$5\times10^{-7}$} &  \bf{-1.0}   & \bf{120} &  \bf{vis}  & 6.9 $\:$ 7.4 & 32 $\:$ 30 & 28 $\:$ 21 & 14 $\:$ 8.6 \\
1.0 & 1.2 & $5\times10^{-7}$ &  -1.0   & 120 &  nir  & 18 $\:$ 19 & 30 $\:$ 26 & 24 $\:$ 15 & 12 $\:$ 5.7 \\
\hline 
   \multicolumn{10}{c}{$dn_p/da \propto a^0$, $k_e=\{0.17,0.27\}$} \\
   \hline 
   \multicolumn{6}{c|}{input parameters}  &  \multicolumn{4}{c}{stellar mass range ($M_{\odot}$)} \\
\hline
$\alpha$ & $\beta$ & $C_0$ & $\gamma_A$ & $\tilde{a}$ & WFS & 0.40-0.95 & 0.95-1.50 & 1.50-2.05 &  2.05-2.60 \\
 \hline
0  &  0  &  $5\times10^{-5}$  &   -1.4   & 35   &  vis  & 0.7 $\:$ 0.1 & 1.3 $\:$ 0.0 & 0.5 $\:$ 0.0 & 0.2 $\:$ 0.0 \\
0  &  1.2 &  $5\times10^{-5}$ &  -1.4   & 35   &  vis  & 1.1 $\:$ 0.2 & 2.1 $\:$ 0.0 & 0.8 $\:$ 0.0 & 0.4 $\:$ 0.0 \\
1.0 & 1.2 & $5\times10^{-5}$ &  -1.4    & 35  &  vis &  0.8 $\:$ 0.2 & 2.4 $\:$ 0.0 & 1.4 $\:$ 0.0 & 0.9 $\:$ 0.0 \\
1.0 & 1.2 & $5\times10^{-5}$ &  -1.4   & 120 &  vis  & 2.7 $\:$ 0.5 & 9.3 $\:$ 0.3 & 5.1 $\:$ 0.0 & 2.7 $\:$ 0.0 \\
\bf{1.0} & \bf{1.2} & \boldmath{$5\times10^{-5}$} &  \bf{-1.0}  & \bf{120} &  \bf{vis}  & 2.7 $\:$ 0.5 & 9.3 $\:$ 0.3 & 8.7 $\:$ 0.0 & 4.6 $\:$ 0.0 \\
1.0 & 1.2 & $5\times10^{-5}$ &  -1.0   & 120 &  nir  & 6.9 $\:$ 2.1 & 8.1 $\:$ 0.2 & 7.1 $\:$ 0.0 & 4.1 $\:$ 0.0 \\
 \hline
0  &  0  &  $5\times10^{-6}$  &   -1.4   & 35   &  vis  & 1.7 $\:$ 0.7 & 3.5 $\:$ 0.3 & 1.2 $\:$ 0.0 & 0.5 $\:$ 0.0   \\
0  &  1.2 &  $5\times10^{-6}$ &  -1.4   & 35   &  vis  & 2.7 $\:$ 1.4 & 6.0 $\:$ 0.6 & 2.3 $\:$ 0.1 & 1.0 $\:$ 0.0 \\
1.0 & 1.2 & $5\times10^{-6}$ &  -1.4    & 35  &  vis &  2.2 $\:$ 0.9 & 7.0 $\:$ 0.7 & 4.0 $\:$ 0.1 & 2.3 $\:$ 0.0 \\
1.0 & 1.2 & $5\times10^{-6}$ &  -1.4   & 120 &  vis  & 3.7 $\:$ 1.7 & 16 $\:$ 3.2 & 8.0 $\:$ 0.7 & 4.2 $\:$ 0.2 \\
\bf{1.0} & \bf{1.2} & \boldmath{$5\times10^{-6}$} &  \bf{-1.0}   & \bf{120} &  \bf{vis}  & 3.7 $\:$ 1.6 & 16 $\:$ 3.1 & 13 $\:$ 0.9 & 6.5 $\:$ 0.2 \\
1.0 & 1.2 & $5\times10^{-6}$ &  -1.0   & 120 &  nir  & 11 $\:$ 6.3 & 15 $\:$ 2.0 & 10 $\:$ 0.2 & 5.6 $\:$ 0.0 \\
\hline
0  &  0  &  $5\times10^{-7}$  &   -1.4   & 35   &  vis  & 3.7 $\:$ 3.9 & 10 $\:$ 8.8 & 4.1 $\:$ 2.7 & 1.6 $\:$ 0.9 \\
0  &  1.2 &  $5\times10^{-7}$ &  -1.4   & 35   &  vis  & 6.4 $\:$ 6.8 & 18 $\:$ 15 & 7.0 $\:$ 4.8 & 2.9 $\:$ 1.6  \\
1.0 & 1.2 & $5\times10^{-7}$ & -1.4   & 35    &  vis &  5.1 $\:$ 5.4 & 21 $\:$ 18 & 12 $\:$ 8.3 & 6.6 $\:$ 3.6 \\
1.0 & 1.2 & $5\times10^{-7}$ &  -1.4   & 120 &  vis  & 8.4 $\:$ 8.7 & 43 $\:$ 41 & 22 $\:$ 18 & 11 $\:$ 7.5 \\
\bf{1.0} & \bf{1.2} & \boldmath{$5\times10^{-7}$} &  \bf{-1.0}   & \bf{120} &  \bf{vis}  & 8.4 $\:$ 8.7 & 43 $\:$ 41 & 33 $\:$ 27 & 16 $\:$ 11 \\
1.0 & 1.2 & $5\times10^{-7}$ &  -1.0   & 120 &  nir  & 23 $\:$ 25 & 40 $\:$ 36 & 28 $\:$ 19 & 14 $\:$ 7.4 \\
\hline
\multicolumn{6}{c|}{Number of Stars Observed with VISWFS} & 87   & 331 & 139 & 53 \\
\multicolumn{6}{c|}{Number of Stars Observed with NIRWFS} & 228 & 340 & 139 & 53 \\ 
\hline
\end{tabular}
\caption{Number of planet detections for a 50 pc survey that targets 610 stars (760 with NIRWFS) with $t_{\tiny{\mbox{age}}}< 1$ Gyr, and $V<9$ ($J<8$). Substantial gains are attained when contrast levels approach $C_0\approx5\times10^{-7}$. Entries for cond03 and fort08 models are shown in the left and right of each column respectively. Two different semimajor axis distributions are considered. The gas-giant planet occurrence rate, $k_e=\{0.17,0.28\}$, is applied to semimajor axis distributions truncated at $\tilde{a}=35$ AU and $\tilde{a}=120$ AU respectively. Calculations allow for multi-planet systems. Simulations with $C_0=5\times10^{-5}$ use a 200 $\lambda/D$ wide FOV. Rows having parameters considered as the baseline case for extrapolation of the RV planet population are bold-faced.}\label{tab:npl_age_bright}
\end{table*}

Of the propitious K and M-stars that are generated in realizations of our age and brightness-selected samples -- those that are young, nearby, and have a planet (e.g., analogous to $\epsilon$ Eridani), only a fraction will have orbit orientations that are conducive to direct detection. While other spectral-types experience similar geometric losses, because many planets are seen in near edge-on orbits and remain hidden behind the coronagraphic mask (e.g., $\beta$ Pic b), this final reduction is enough to essentially preclude planet detections around M-stars unless a NIRWFS is employed. To illustrate the difficulty in finding viable low-mass targets, we count those available from \citet{zuckerman_song_04} and \citet{shkolnik_09}. The total number of single K and M-stars with $V<9$ and $s\leq50$ pc, including the $\beta$ Pictoris, TW Hydrae, Tucana-Horologium, $\eta$ Cha, and AB Dor groups, is only 16. With planet occurrence rates $\lesssim10$\% at wide separations, it is difficult to make a substantial number of detections with such small samples. The GPI will use a wavefront sensor\footnote{Most ``extreme" AO systems will employ two wavefront sensors: one that is sensitive to dynamic wavefront errors  and runs at a frequency $f$$\approx$1,000 Hz, and another that is more sensitive to static errors, running at a frequency of $f$$\approx$1 Hz \citep{wallace_10}. Here we are referring to the dynamic wavefront sensor.} that operates in the I-band and can therefore expect a number of planet detections that lies in between the (brown dash and red dash-dot) curves shown in Figs. 4, 5.

A scenario in which correlations between star mass and planet occurrence rate are weak can improve the number of detections at low star masses. Raising the overall planet occurrence rate can also help, although the size of the error-bar on $k$ from Eqn.~\ref{eqn:mass_met} is already small. Previous high-contrast imaging observations have placed firm upper-limits on the overall frequency of massive ($m_p\geq4M_J$) planets in wide orbits at $20\%$ \citep{nielsen_10}. Otherwise, loosening age and brightness selection cuts can enhance the number of planets imaged around low-mass stars at the expense of detection efficiency. Results from the volume-limited case (Fig. 2, 3) show the number and rate of detections in the limit as both criteria are completely removed. 

The distribution of planet detections around massive stars is similar to that of a volume-limited survey, indicating that a significant fraction of stars with $M_*\gtrsim1.8M_{\odot}$ have made the cut because they are intrinsically young and bright, owing to the applied selection bias. Fig. 6 shows the relationship between the median and maximum age of stars used for our simulations. Given the difficulty in age-dating stars in the field, stellar mass serves as an excellent proxy and can aid in the selection of promising high-contrast imaging targets for a number of reasons.

\begin{figure}[!h]
\begin{center}
\includegraphics[height=2.4in]{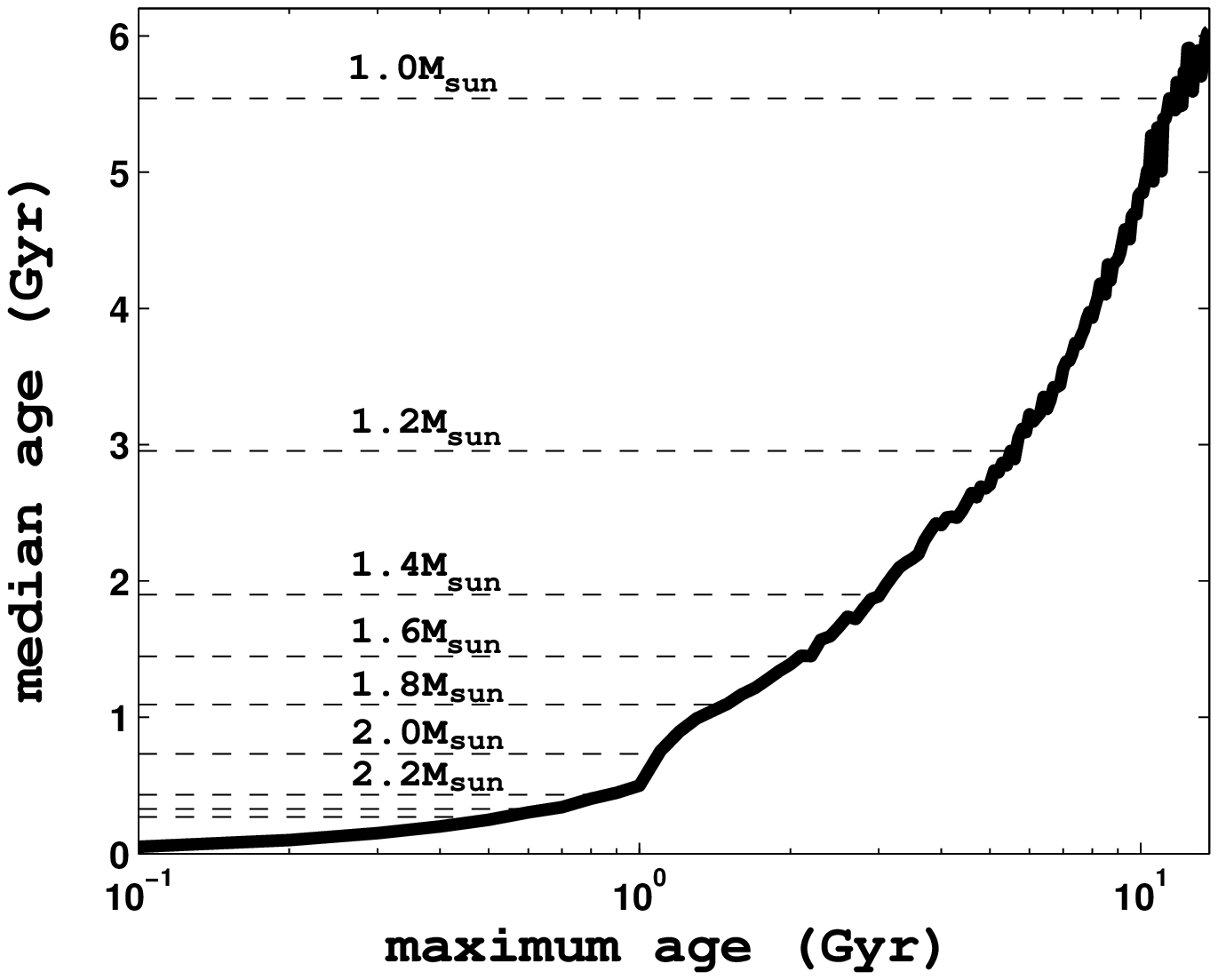}
\caption{Median and maximum ages used for Monte Carlo simulations to ensure consistency with (pre-)main-sequence or subgiant branch evolution. Intrinsically young, massive stars have bright planets compared to low-mass stars, increasing the likelihood for a direct detection. In the absence of information regarding membership to a moving group or association, stellar mass may be used as a trivial aid in the selection of viable field stars for high-contrast imaging targets.}
\end{center}\label{fig:intrinsic_youth}
\end{figure}

\subsubsection{Planet Detection Rates}
Dividing the number of planet detections in each stellar mass bin by the number of targets observed in each stellar mass bin, we find that the probability for imaging a planet is enhanced for all spectral-types compared to the volume-limited case (as expected), and that massive stars are again preferred targets (Fig. 4, 5). Both models indicate however that this result is sensitive to the value of $\alpha$, the power-law index that controls the planet occurrence rate as a function of stellar mass. When $\alpha=0$, the detection rate becomes relatively flat in the range, $1.0 \leq M_*/M_{\odot} \leq 2.6$, and exhibits a small negative slope (i.e., efficiencies favoring lower mass stars) for $M_*<1.0M_{\odot}$. Although the slope changes sign in a regime where few planets are imaged, a NIRWFS can provide access to a sufficient number of stars to notice such a trend in practice when $C_0\approx5\times10^{-7}$. Correlations between metallicity and planet occurrence rate can further improve number statistics at low star masses. Dropping our assumption that semimajor axis extent scales with stellar mass yields slightly smaller (more negative) slopes (see $\S$\ref{sec:tuc}).

Given these results, we conclude that it will be important to carefully examine the detection rate of future high-contrast imaging programs as a function of stellar mass by making plots similar to Fig. 4, 5. Such diagnostics require a large sample of targets but can potentially help to discriminate between planet formation scenarios. \emph{When targets are chosen carefully, so as to minimize the effects that stellar age and brightness have on detection efficiencies, the dependence of the planet occurrence rate upon star mass (Equ. 1) is strong enough to help assess whether the processes responsible for forming the RV planet population ($a<6$ AU) may also operate at larger separations.} 

Planet evolutionary models are sensitive to age, and stellar ages are notoriously difficult to determine with accuracy \citep{soderblom_10}. We find that our results are robust to perturbations to the distribution in Fig. 1, provided that there remains some semblance of a roughly constant star-formation rate over the past 4-6 Gyr \citep{mamajek_hillenbrand_08}. 

\subsection{Current Technology}\label{sec:C05}
Motivated by the direct detection of planets orbiting $\beta$ Pictoris (A6V) \citep{lagrange_10}, HR 8799 (A5V) \citep{marois_08,marois_10}, and Fomalhaut\footnote{Fomalhaut b has mysterious colors and has yet to be detected from the ground in the NIR.}  (A3V) \citep{kalas_08}, we have performed additional calculations using contrast levels commensurate with those achieved by instruments using standard AO, to compare our results with recent observations. Repeating the age-and-brightness selected survey where targets within 50 pc are chosen based on their youth, $t_{\mbox{\tiny{age}}}<1$ Gyr, and brightness for AO correction, $V<9$ ($J<8$), Fig. 7 shows our results for the number and rate of planet detections using the cond03 models when setting contrast levels to $C_0=5\times10^{-5}$ and expanding the search region to $\pm100 \;\lambda / D$ ($11\prime\prime$ wide) from the star to better reflect the typical FOV of current instruments \citep{marois_08,biller_10}. The number of planet detections using both planet evolutionary models and various input parameters is also shown in Table 3. 

\begin{figure*}[!t]
\begin{center}
\includegraphics[height=2.4in]{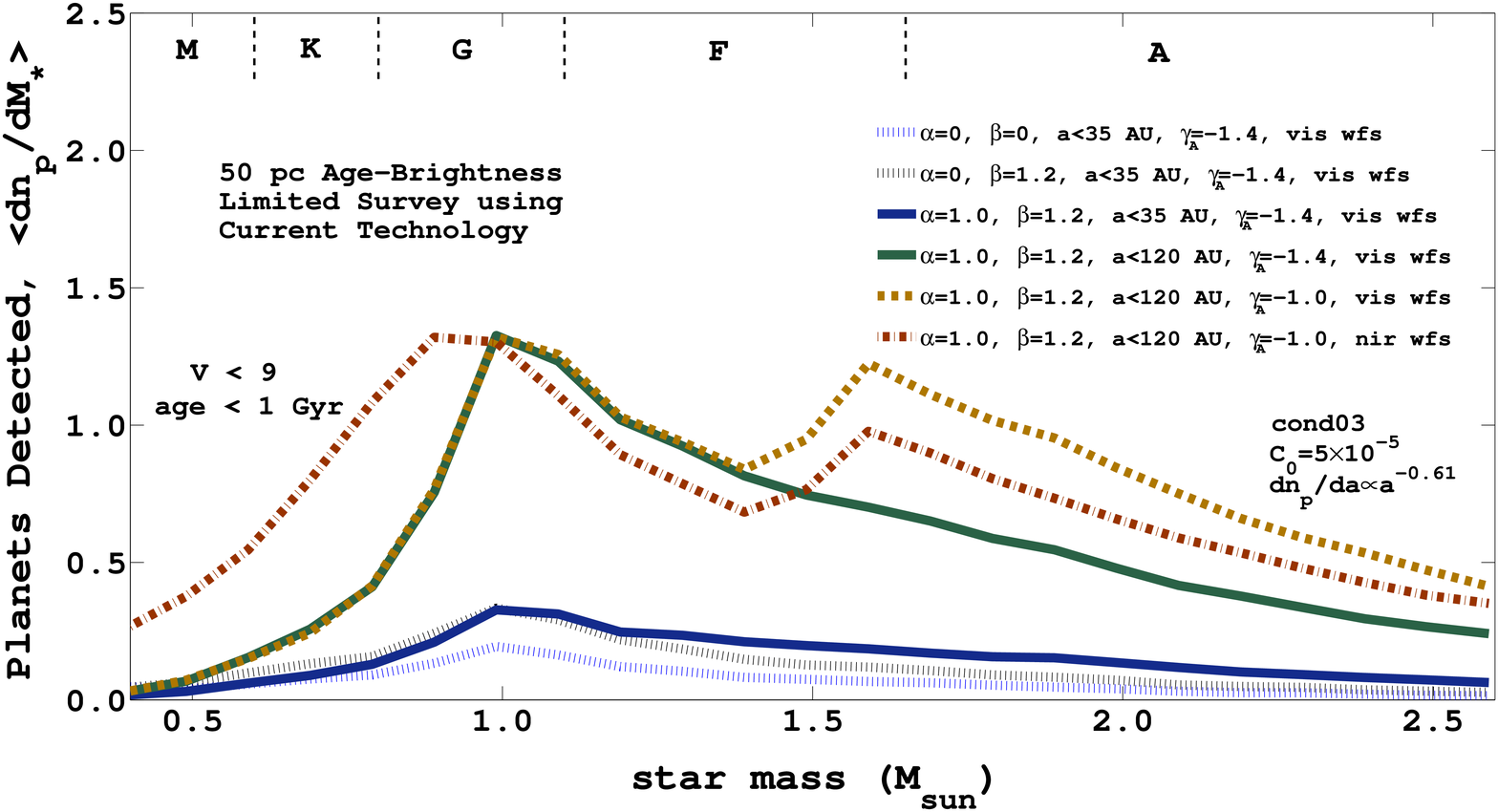}
\includegraphics[height=2.4in]{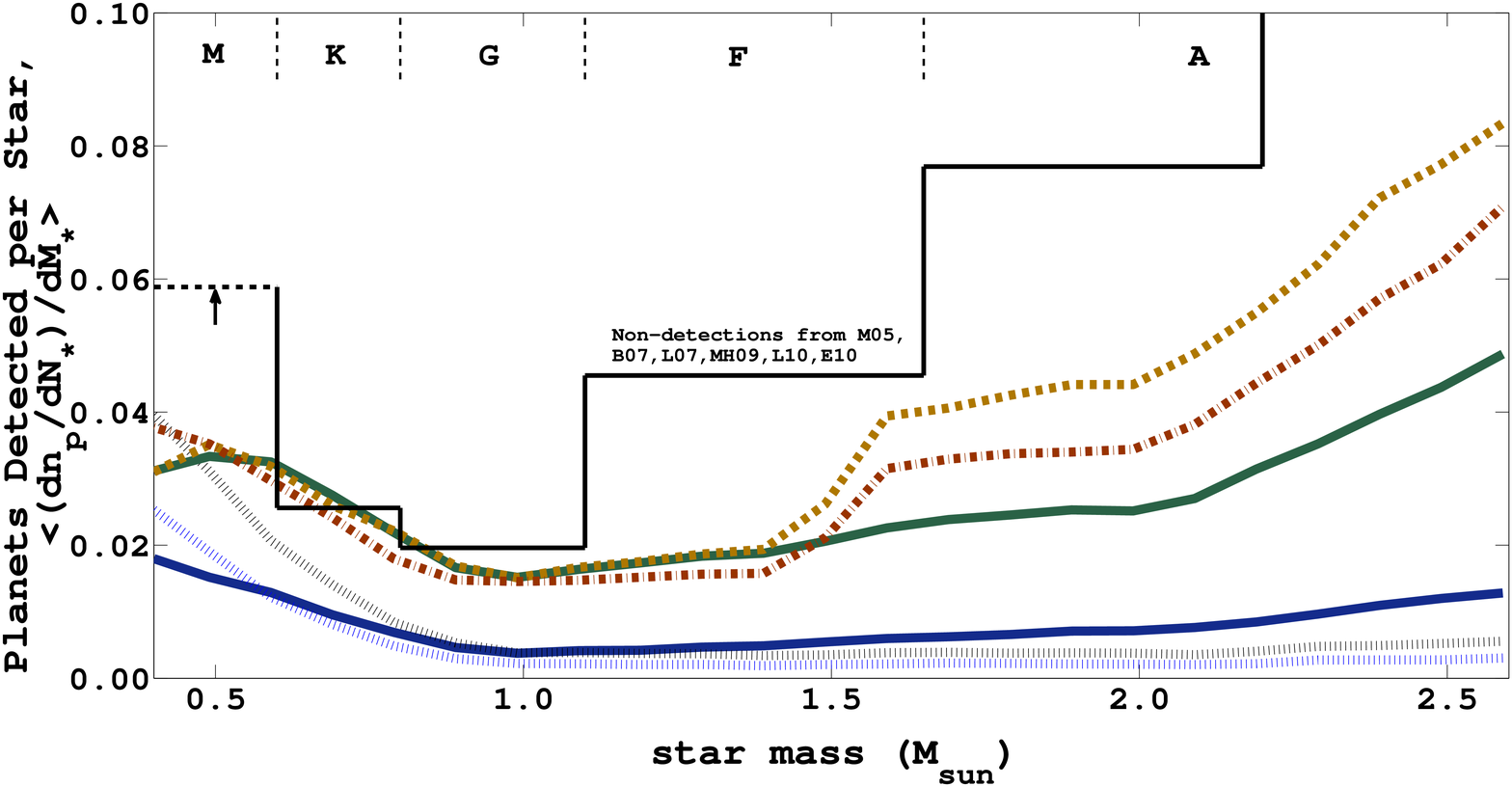}
\caption{Planet detection numbers and rates as a function of stellar mass for a 50 pc age-and-brightness selected ground-based survey using contrast levels of $C_0=5\times10^{-5}$, comparable to present-day instruments. Low-mass stars yield few detections and high-mass stars offer the highest efficiency when planets at wide separations resemble the RV sample. For reference, 95\% confidence levels from a number of high-contrast surveys yielding non-detections ($V<9$, $t_{\mbox{\tiny{age}}}<1$ Gyr) are over-plotted in the bottom panel. Constraints for M-stars represent a lower-limit since we have included $V<11$ targets in this bin. Our calculations are thus consistent with high-contrast imaging surveys conducted to date in terms of both relative and absolute numbers. Simulation results may be scaled up or down based on the actual occurrence rate and truncation radius of planets in wide orbits (e.g., consider the difference between the solid blue and solid green curves).}
\end{center}\label{fig:curr_tech}
\end{figure*}

We find that degrading contrast levels yields results that resemble Figs. 4, 5, only the number and efficiency of detections is essentially scaled down by a factor of several for all curves. Careful examination between the two cases shows that low-mass stars have gained some relative ground when $C_0$ is adjusted from $5\times10^{-7}$ to $5\times10^{-5}$, on account of less demanding contrast requirements compared to massive stars, but nevertheless the results are qualitatively similar. Low-mass stars generate few detections, though are relatively efficient targets when $\alpha=0$; the detection rate curve is flat beyond $M_*=0.9M_{\odot}$ when $\alpha=0$; and high-mass stars yield a comparable number of detections as low-mass stars but with high detection rates when $\alpha=1$, particularly when the power-law index governing planet masses is modified from $\gamma_A=-1.4$ to $\gamma_A=-1.0$. 

In other words, setting $C_0=5\times10^{-5}$ we find that naive extrapolation of the known planet population, i.e., those discovered by the RV technique, to separations wider than several AU is consistent with A-stars dominating the first planet imaging discoveries (see also $\S$\ref{sec:tuc}). This interpretation of our simulations is reinforced by the null-results of other high-contrast programs that have targeted nearby solar-type stars: \citet{masciadri_05} (M05), \citet{biller_07} (B07), \citet{lafreniere_07} (L07), and \citet{metchev_hillenbrand_09} (deep sample only) (MH09) selected targets based primarily on proximity and age and were marginally sensitive to wide-separation planets, with $C_0\approx10^{-4}$-$10^{-5}$. Combined, these surveys observed: 50 K-stars, 61 G-stars, 16 F-stars, but only 1 A-star within 50 pc ($V<9$, $t_{\mbox{\tiny{age}}}<1$ Gyr).\footnote{Consult \cite{nielsen_10} for more details.} 

\begin{figure*}[!t]
\begin{center}
\includegraphics[height=2.4in]{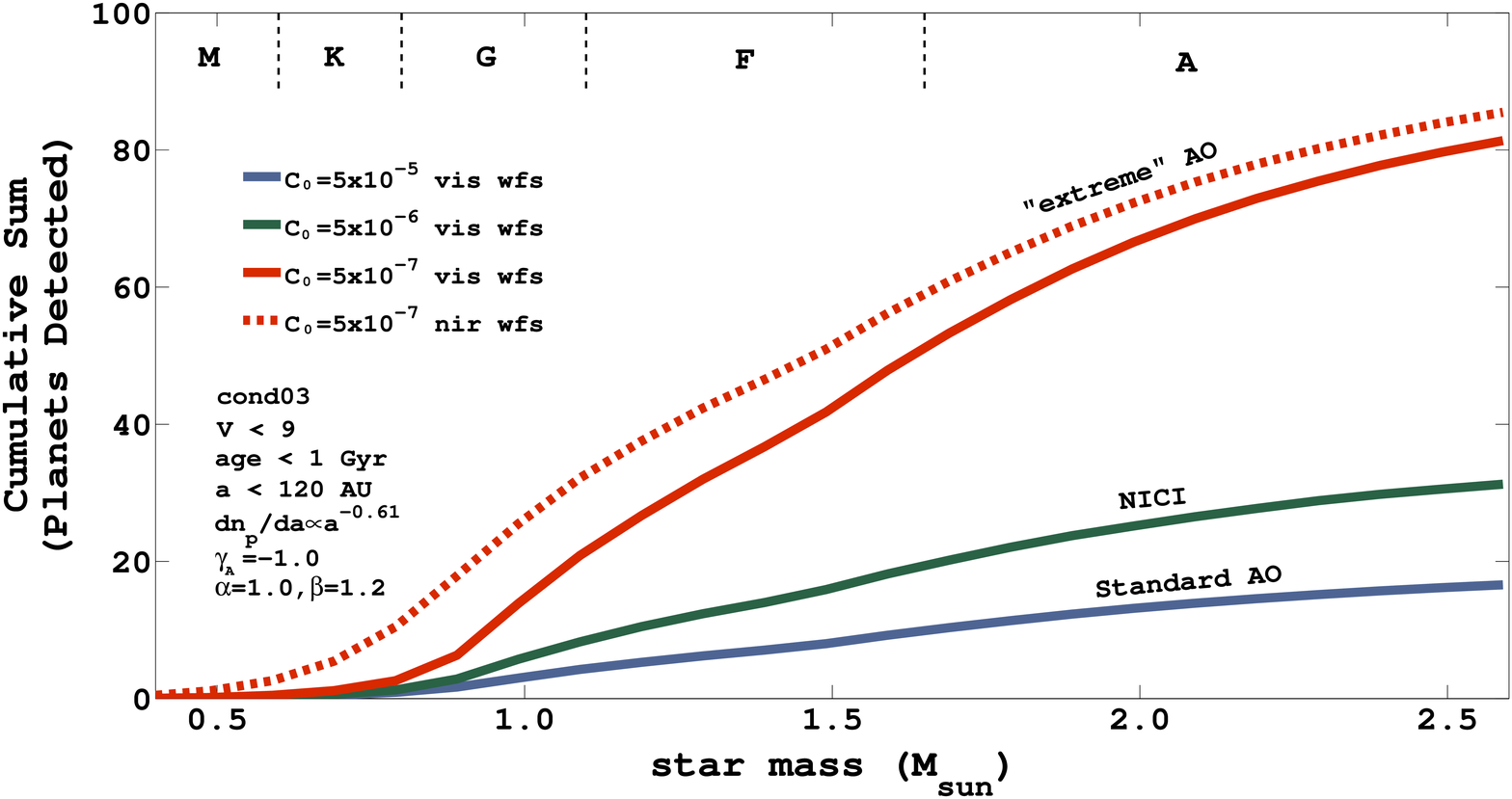}
\includegraphics[height=2.4in]{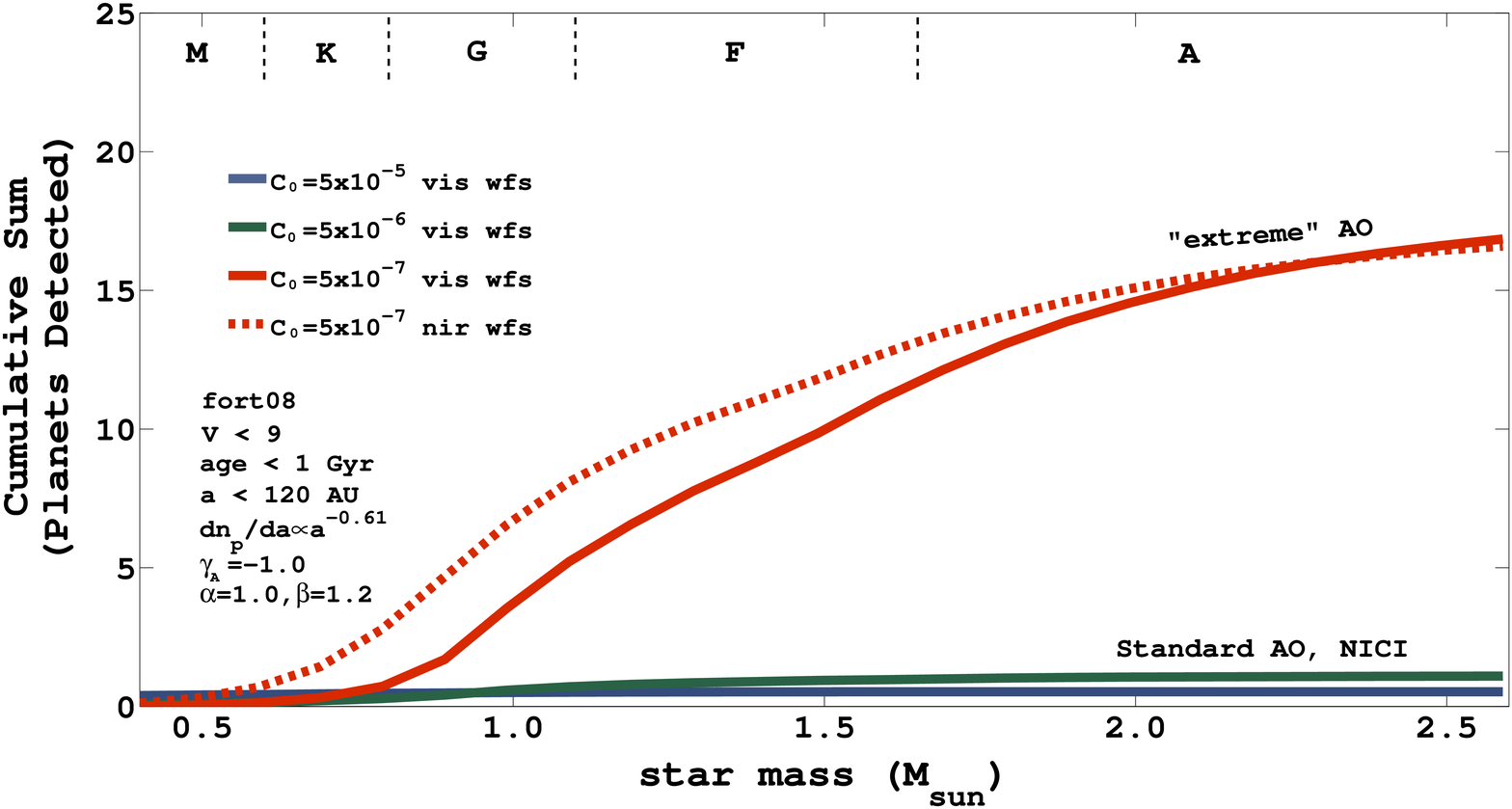}
\caption{Cumulative number of planets detected as a function of stellar mass for a 50 pc age-and-brightness selected survey using the cond03 models (top) and fort08 models (bottom). Several characteristic contrast levels are considered. The number of detections using current technology well-matches the results from observational surveys to date, predicting between 0.6-16.6 planet imaging discoveries for $C_0=5\times10^{-5}$. Surveys resembling the NICI campaign can at most double the number of planet imaging detections. Substantial gains occur when contrast levels improve beyond $C_0=5\times10^{-6}$ and low-mass planets ($\approx$$0.5M_J$) are imaged.}
\end{center}\label{fig:cumsum}
\end{figure*}

The combination of null results from these early high-contrast imaging surveys along with the detection of massive planets with wide orbits around subgiant stars (former or ``retired" A-stars) using the RV technique has prompted other groups to focus more-so on massive stars. Recently, \citet{leconte_10} (L10) and \citet{ehrenreich_10} (E10) have published non-detections for substellar companions orbiting: 3 G-stars, 14 F-stars, 22 A-stars, and 4 B-stars within 60 pc ($V<9$, $t_{\mbox{\tiny{age}}}<1$ Gyr), achieving similar contrast levels. As shown in Fig. 7, the average number of young, nearby stars required to detect a large-separation planet, given our assumptions and no other previous knowledge on the presence of companions or a debris disk\footnote{HR 8799, $\beta$ Pictoris, and Fomalhaut are clearly special cases in this regard, each having prominent debris disks.}, is higher than that reported for $\tilde{a}<120$ AU ($95\%$ confidence level). 

Using parameter values $\alpha=1$, $\beta=1.2$, $\tilde{a}=120$ AU, $\gamma_A=-1.0$, and $dn_p/da \propto a^{-0.61}$ with a VISWFS, the fort08 and cond03 models yield $0.6\pm0.2$ and $16.6\pm4.4$ total planet detections (ensemble average), taking into account the uncertainty in parameters from Equ. 1 ($k_e=0.27\pm0.04$, $\alpha=1.0\pm0.3$, $\beta=1.2\pm0.2$). The number of directly imaged planets orbiting young stars within 50 pc currently stands at 5 (HR 8799 bcde, $\beta$ Pic b). Our simulations using the ``core-accretion" and ``hot-start" thermal evolutionary models, which represent extrema in brightness predictions for young planets (see $\S$\ref{sec:planets}), are thus in excellent agreement with observational results to date in terms of both relative and absolute detection rates, straddling the number of known planets with expectations for several more discoveries in the near-term. 

Fig. 8 shows the cumulative number of planet detections as a function of stellar mass for various contrast levels using the cond03 models and same input values listed above. Both a VISWFS and NIRWFS are considered for next-generation instruments. The Near-Infrared Coronagraphic Imager (NICI) campaign occupies an interesting parameter space and is conducting a thorough search of $\sim300$ young stars using the Gemini S. AO system, a coronagraph, and speckle suppression techniques that generate unprecedented contrast levels \citep{liu_10_spie,biller_10,wahhaj_11}. Despite these advances in hardware and data-processing, routinely achieving sensitivity to $>1M_J$ companions, our simulations indicate that NICI and other surveys of similar scope will at most double the number of planet images. There exist solar-type stars within 50 pc with imageable planets at $C_0\approx5\times10^{-6}$, but the efficiency of observations in this regime is low, requiring a dedicated effort involving observations of all available nearby young FGK stars, particularly if planet semimajor axes extend only to several tens of AU. 

We predict that substantial gains will be obtained only when contrast levels improve by two orders of magnitude compared to observations using standard AO. Instruments that employ ``extreme" AO systems will increase the number of planets imaged directly around $V<9$ ($J<8$) stars within 50 pc, yielding between 12-81 discoveries total. Observations of fainter stars can further boost the number of detections at the expense of many more nights at the observatory ($\S$\ref{sec:volume}). 

\begin{figure*}[!t]
\begin{center}
\includegraphics[height=3.0in]{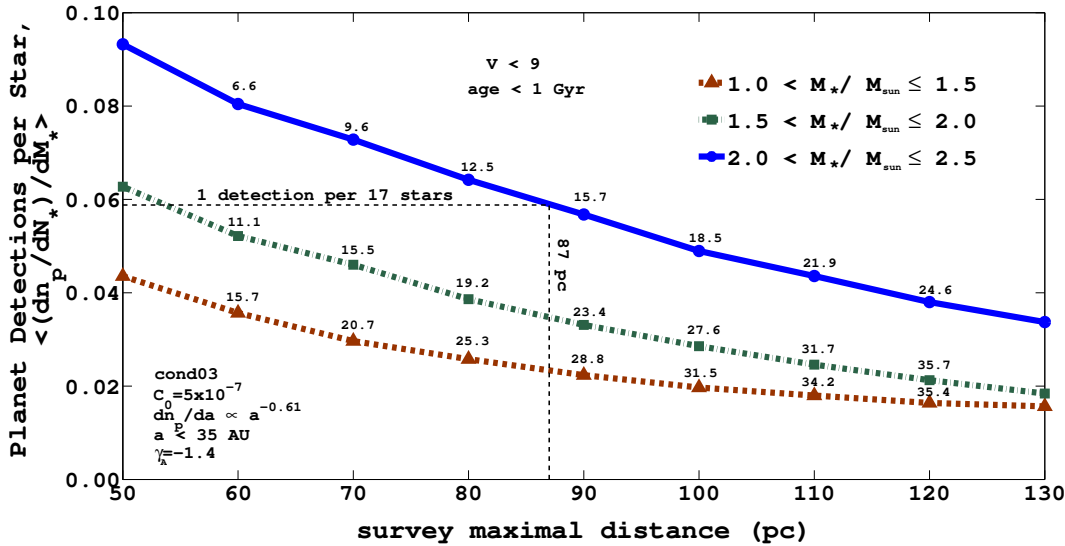}
\caption{Monte Carlo simulations of an age-and-brightness limited high-contrast imaging survey targeting stars with incrementally larger maximum distances. Such observations can provide an additional factor of at least $2-3$ in the total number of planet detections for $M_*>1.0M_{\odot}$ stars compared to canonical surveys restricted to 50 pc. Massive stars yield the best observing efficiency, offering more than one detection per 17 targets for stars out to a distance of $\approx 87$ pc averaged over the entire volume. Detection rates approach a minimum as the survey becomes magnitude-limited due to the applied $V<9$ target selection cut. The total cumulative number of detections is shown above each data point. For example, the number of detections between $60-70$ pc for $1.5\leq M_* / M_{\odot} < 2.0$. stars is $15.5-11.1=4.4$ planets. Simulations assume a sky background limit of 22 mag, semimajor axis distribution $dn_p/da \propto a^{-0.61}$, with $a<35$ AU extent, and planet mass power-law index $\gamma_A=-1.4$.}
\end{center}\label{fig:dist}
\end{figure*}

\subsection{More Distant Stars}\label{sec:dist}
Several dozen extrasolar planets orbiting stars within 50 pc of the Sun will be imaged and characterized by high-contrast instruments over the next few years. While preliminary trends may begin to emerge from these datasets (e.g., see \citet{marcy_99} for a review of the RV population when it comprised 17 exoplanets), it is important to collect the largest possible sample for statistical analyses to inform our understanding of gas-giant planet formation and evolution. Indeed, planet properties and correlations with host star properties are likely to involve distributions that are multi-dimensional, multi-modal, and possibly time-dependent at the youngest ages. 

We have performed calculations for young and bright stars beyond 50 pc to gain a handle on the expected yield of magnitude-limited high-contrast surveys. Results for the planet detection rate setting $C_0=5\times10^{-7}$ are shown in Fig. 9 for stars separated into three different mass bins, $1.0\leq M_*/M_{\odot}<1.5$,  $1.5\leq M_*/M_{\odot}<2.0$, and  $2.0\leq M_*/M_{\odot}<2.5$, along with the cumulative number of detections as the survey maximum distance is increased outwards in 10 pc increments. Targets are again chosen such that $V<9$ ($J<8$ with NIRWFS) and $t_{\tiny{\mbox{age}}}<1$ Gyr using the distributions from Table 1. Simulation input is selected to represent a ``pessimistic" case in order to place a lower-limit on the number of planet detections gained. We have assumed that the semimajor axis distribution of the RV planet population extends only to $\tilde{a}=35$ AU, and that the sky background is bright, limiting planet detections to apparent magnitudes $m_{\mbox{\tiny{JHK}}}<22$. Further, we have removed the assumption that more massive planets orbit stars with mass $M_*\geq1.5M_{\odot}$. 

Despite assuming non-ideal input parameters, we find that additional wide-separation companions may be imaged around $1.0 < M_*/M_{\odot} < 2.5$ stars well beyond 50 pc, the canonical distance that most planned surveys are restricted.  Consistent with previous sections, the most massive stars ($M_*>2.0M_{\odot}$) offer the highest detection rates. Average efficiencies for each group fall slowly with increasing distance and asymptotically approach a minimum value as the maximum number of available stars in each mass bin is reached from the applied $V<9$ brightness selection cut. The number of detections does not plateau until distances exceeding $100$ pc. 


Stars with mass, $2.0<M_*/M_{\odot}\leq2.5$, offer detection efficiencies of at least one planet detection per 17 targets out to a distance of 87 pc (averaged over the entire volume). For reference, approximately 10 stars per good-weather-night may be observed with an 8-10m telescope using the ADI technique. The number of planets detected in each stellar mass bin increases from 11.3, 7.6, and 4.4 at 50 pc to 25.3, 19.2, and 12.5 at 80 pc respectively, suggesting that a factor of 2-3 gain in the wide-separation planet population is possible by including more distant targets. We find that improving the IWA from 4$\; \lambda / D$ to $3 \; \lambda / D$ can improve the number of detections by another $\approx$$11\%$ in each mass bin over a similar space volume. Although proper motions in this regime are small, requiring long time baselines for follow-up astrometry\footnote{Candidates have been confirmed as common proper-motion companions around stars as distant as $\approx160$ pc \citep{lafreniere_11}. Further, it may be possible to establish association indirectly using spectroscopic characterization.}, augmenting target lists by including stars with distances between 50 pc and the nearest star-forming regions at $\approx$140 pc will maximize the overall science output of future instruments. 

\subsection{A Nearby Association}\label{sec:tuc}
Members of nearby moving groups and associations fall under the category of target selection by age because the constituent members are likely to have formed around the same time. When a cluster of young stars are also situated at roughly the same distance from the Sun, they represent a special case in high-contrast imaging and convenient control for comparing the number and rate of planet detections to the volume-limited and age-and-brightness-selected surveys presented previously. In this section, we further elucidate our results by simulating observations of a collection of nearby, young stars, each with an age, $t_{\tiny{\mbox{age}}}=30\pm10$ Myr, at a distance, $s=48\pm7$ pc. Such a consortium may be considered as representative of the Tucana-Horologium association \citep{torres_08}. 

\begin{figure*}[!t]
\begin{center}
\includegraphics[height=2.4in]{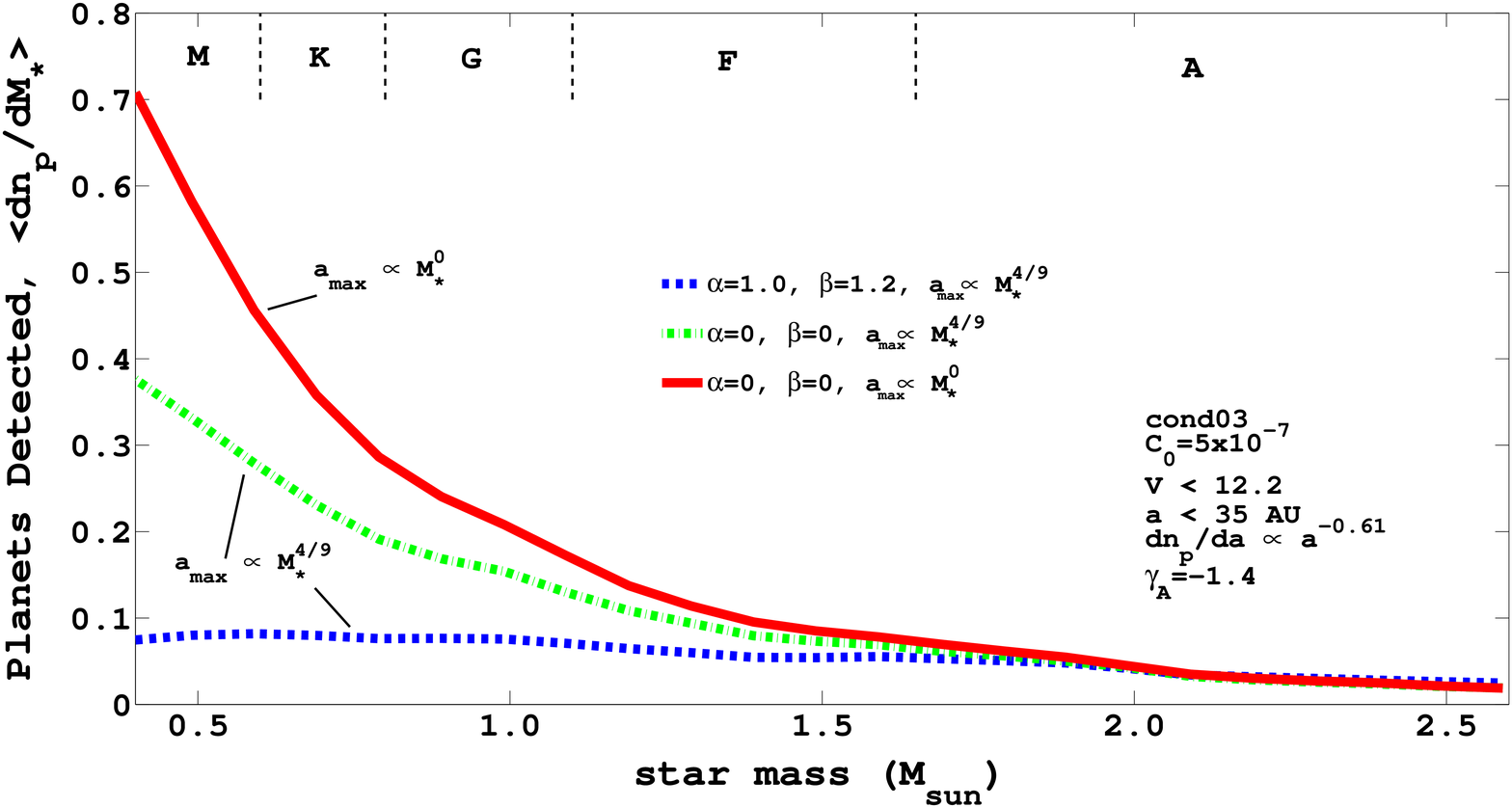} 
\includegraphics[height=2.4in]{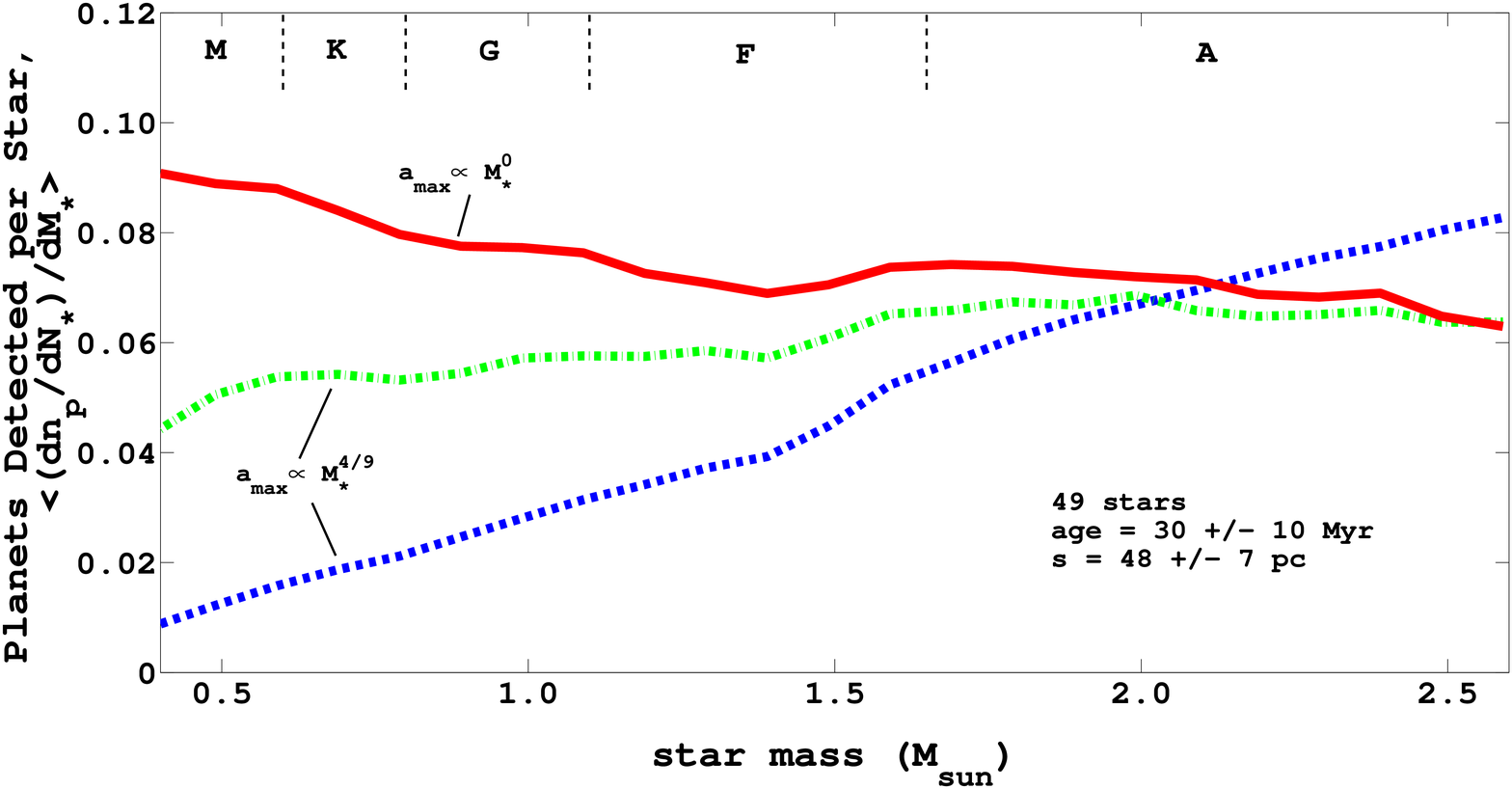}
\caption{Monte Carlo simulations of a nearby young stellar association synthesized to resemble Tucana-Horologium, using $C_0=5\times10^{-7}$, $\tilde{a}=35$ AU, and the cond03 planet evolutionary models. Gas-giant planet occurrence-rates as a function of stellar mass, $\alpha$, and metallicity, $\beta$, are changed, along with the scaling of the size of the planet formation zone with stellar mass. Low-mass stars dominate the number of detections as a result of the common age and distance shared by constituent members. The detection rate rises towards higher stellar masses when correlations between star mass and planet properties are considered. The slope of detection rate curve is negative when assumptions about planet occurrence rate are dropped and we relax our assertion that the maximal extent of planet orbits scales with stellar mass. In this case, low-mass stars can yield the most detections and highest detection efficiencies. The value of detection rate slopes may be used as a diagnostic to help discriminate between planet formation scenarios.}
\end{center}\label{fig:loose}
\end{figure*}

We consider three different cases using $\tilde{a}=35$ AU, $C_0=5\times10^{-7}$, and a \citet{miller_scalo_79} IMF:  
\begin{description}
\item[(I)] $\alpha=1$, $\beta=1.2$, $a_{\mbox{\tiny{max}}}=(M_*/2.5M_{\odot})^{4/9}\;\tilde{a}$
\item[(II)] $\alpha=0$, $\beta=0$, $a_{\mbox{\tiny{max}}}=(M_*/2.5M_{\odot})^{4/9}\;\tilde{a}$ 
\item[(III)] $\alpha=0$, $\beta=0$, $a_{\mbox{\tiny{max}}}=(M_*/2.5M_{\odot})^0\;\tilde{a}$  
\end{description}
The first two cases were treated in the 50 pc volume-limited and age-and-brightness-selected surveys. In the third case, we relax our assumption that the planet formation zone grows with stellar mass by setting the maximal extent of semimajor axes $a_{\mbox{\tiny{max}}}=35$ AU for all targets. For each scenario, the above age and distance ranges are used as limit values for uniform sampling. Tucana-Horologium currently has 49 known members \citep{zuckerman_song_04}. We simulate observations of all members ($V<12.2$) and construct $10^4$ ensembles, a larger number than in previous sections, to reduce the statistical noise associated with the small number of targets. The metallicity of each star is set to $\mbox{[Fe/H]}=-0.25$ dex \citep{makarov_07b}. Only one planet evolutionary model is considered, because different models yield scaled versions of one another when the age is fixed; i.e., the number of planets detected for different models is related by a multiplicative factor. Fig. 10 shows our results assuming ``hot-start" initial conditions. 

We find that low-mass stars yield the most planet detections and that there is a systematic decline with increasing stellar mass. This simple dependence is a result of the features held in common with all stars in the association. With the same approximate age and distance, the number of planet detections is governed primarily by the stellar IMF. Setting $\alpha=0$ and $\beta=0$ changes the slope of the planet detection curve further in favor of low-mass stars. The steepest slope occurs when $\alpha=0$, $\beta=0$, and $a_{\mbox{\tiny{max}}}=(M_*/2.5M_{\odot})^{0}\;\tilde{a}$. In this case special case, the ratio of the number of planet detections for high-mass stars to that of low-mass stars is close to the ratio of stars from the IMF -- for instance, $n_p(2.0M_{\odot})/n_p(1.0M_{\odot})=0.24$ compared to $(2.0M_{\odot}/1.0M_{\odot})^{-2}=0.25$ -- though slightly smaller because massive stars command a deeper contrast ratio. This is effectively a verification of our code. Indeed, controlling for stellar brightness and ``turning off" the effects from the AO system produces a slope in the number of detections that matches exactly that dictated by the IMF.

Dividing the number of planet detections by the number of targets observed in each stellar mass bin, we find that massive stars are preferred targets when occurrence-rate and semimajor axis correlations are applied. However, setting $\alpha=0$ (changing $\beta$ only creates a constant offset) and eliminating the correlation between $a_{\mbox{\tiny{max}}}$ and $M_*$ results in an average slope of $-0.07$ dex / $M_{\odot}$ over the stellar mass range considered. \emph{This simplest case is the only one, including simulations from all previous sections, where M-stars both dominate the number of detections and have the highest observational efficiency.}

\begin{table*}[!t]
  \centering 
   \begin{tabular}{lccccccccc}
   \multicolumn{8}{c}{} \\
   \hline    
Star                   &  SpTy   &   $M_* / M_{\odot}$  & $m_p / M_J$   &   $V$     &  Assoc.       & Age (Myr) &  a (AU)  &  $s$ (pc)  &   $\rho/\bar{d}$     \\
 \hline
 Fomalhaut       &      A3   &   2.1                        & $<$3                 &  1.2             &  Castor                &      200            &    119               &   $7.7$      &        0.5    \\
 $\beta$ Pic      &      A6   &   1.8                        &  $8^{+5}_{-2}$  &  3.9              &   $\beta$ Pic        &       10-12       &    8-15              &  $19.3$    &         0.7    \\
 HR 8799          &      A5   &   1.6                        &  5-10                &  6.0              &    Columba?          &    20-150       &   15-68             &  $39.4$    &         1.8     \\
\hline
 AB   Pic           &    K2     &  0.8                           &  $13.5\pm0.5$   &  9.2                &   Tuc-Hor      &      30        &    260           &  $47$      &    0.2     \\
GQ Lup           &     K7     &  0.7                           &  $21.5\pm20.5$  & 11.4               &  Lupus          &        1       &     103           & $140$      &   $0.2$  \\
1RXS J1609    &     K7    &   0.7                          &  $\approx$8         &  $I$=11.0        &   USco        &         2-5     &     330           &  $145$     &   0.1      \\
 GSC 06214    &     M1    &  0.6                          &    $13.5\pm2.5$    &  $I$=11.1         &    USco        &     2-5         &   320             &     145      &  0.1       \\
\hline 
   \end{tabular}
\caption{Directly imaged planetary-mass ($m_p<15M_J$) companions with stellar primaries as of February 25, 2011 (http://exoplanet.eu/). There currently exists an apparent dichotomy between the detection of planets around A-stars in the solar neighborhood and the detection of ultra-wide separation planets around late-type stars in clusters. This division may be roughly parameterized according to the ratio of the cluster size to the average distance of its members from the Sun, $\rho/\bar{d}$. Our calculations show that massive stars are preferred targets when searching within the solar neighborhood ($s<50$ pc), provided strong correlations between star mass and planet properties are present. Young stellar clusters represent a special case in high-contrast imaging and also suffer from small number statistics, yielding the most planet detections around MK stars, irrespective of such correlations. If ultra-wide separation planets form by a separate process than those orbiting close to their star, the planet detection rate for clusters will have a negative or approximately flat slope when plotted against stellar mass. This diagnostic may be used to help differentiate between formation scenarios. Numbers used for ages, distances, and spatial scatter are from \cite{beichman_10}, \cite{makarov_07}, \cite{barrado_98}, the Extrasolar Planets Encyclopedia, and NStED.}
\label{tab:images}
\end{table*}

The vast majority of Tucana-Horologium members have been searched previously with AO and a coronagraph. There is one companion with a mass at the planet/BD boundary that was detected directly orbiting the K2V star AB Pic ($V=9.2$) \citep{chauvin_05_abpic}. Performing simulations that mimic present-day observations, by degrading contrast levels to $C_0=5\times10^{-5}$, increasing the FOV (see $\S$\ref{sec:C05} for details), and also setting $\tilde{a}=300$ AU, we find that the three cases yield a cumulative number of $0.3\pm0.1$, $1.2\pm0.1$, and $1.5\pm0.1$ imaged planets over the MK spectral-type range respectively. 

The Tucana-Horologium association is currently incomplete at low stellar masses, because the faintest members have yet to be identified. One might therefore cautiously speculate that the detection of a single planet candidate thus far favors case III. However, our simulations yield the same answer for cases II, III to within rounding errors. Also, a different formation mechanism could produce a different overall planet occurrence rate and we have set $k_e=0.17$ for each case. Further, planet evolutionary models are too uncertain at this age to justify using calculations involving absolute numbers alone to discriminate between possible formation scenarios.

These results are consistent with the recent detection of large-separation, planetary-mass objects found orbiting late-type members of moving groups, associations, and star-forming regions. Table 4 shows a compilation of directly imaged planets ($m_p<15M_J$) to date according to the exoplanet encyclopedia (http://exoplanet.eu/). There currently exists a dichotomy whereby the host stars fall cleanly into two categories of nearby massive stars on one hand, and low-mass stars in more distant associations/clusters on the other hand. 

The preference for planet detections around low-mass stars may also be explained by an observational bias resulting from small number statistics. For a given tightly bound cluster, there are generally too few members to have a reasonable chance of imaging a planet around a high-mass star, despite correlations that may exist between star mass and planet properties. For example, Tucana-Horologium is one of the largest nearby associations and contains only 4 A-stars. This bias is exacerbated by the fact that: (1) half of cluster members are binaries, and high-contrast programs nominally avoid binaries; (2) most stellar clusters are located at 140 pc, effectively reducing the overall planet occurrence rate to its value at large orbital separations; (3) current estimates of the frequency of ultra-wide separation planetary-mass companions are $\approx$1-4\% \citep{metchev_hillenbrand_09,ireland_11}.

We find that the two host star groups, late-type and early-type, may be roughly separated according to $\rho/\bar{d}$, the ratio of the characteristic cluster size (scatter) to the mean distance of cluster members from the Sun. When $\rho/\bar{d}$ is large and nearby targets comprised of different moving groups, associations, recently dispersed clusters, and stars that have formed in relative isolation -- each having different ages -- are spatially mixed (often labeled as field stars beyond a given age or when exhibiting space motion that is unreconcilable with known kinematic groups), the resulting number and rate of detections follow the trends in Fig. 4, 5, and 7. When $\rho/\bar{d}$ is small and a group of tightly clustered stars that unambiguously share the same age and distance are observed, the resulting number and rate of detections follow the trends in Fig. 10. 


A change in the sign of the detection rate slope is a specific prediction that can be used to help differentiate between formation scenarios (Fig. 10). If a negative detection rate slope is found for wide separation planets orbiting MK stars in tight clusters, then these planets could not have formed in the same way as HR 8799 bcde or $\beta$ Pic b, because we find that strong correlations between star mass and planet properties are required to account for their existence ($\S$\ref{sec:C05}). While measurement of a positive detection rate slope is an inconclusive result by itself, measurement of the slope value at the 10\% level can offer a more definitive discriminatory test. Diagnostics based on stellar mass can complement other studies, such as those involving host star metallicities, semimajor axis distributions, and inferences made from absolute detection rates.  Several teams are now conducting dedicated surveys of star-forming regions \citep{todorov_10,ireland_11,lafreniere_11}. Their results can help determine whether multiple formation channels are at work. 

\section{Caveats}
In this section, we list important caveats and assumptions that have gone into the paper and describe their qualitative influence on the results: \\

{\noindent}(1) While this study is concerned primarily with calculations of the relative behavior in the number and efficiency of planet detections as a function of stellar mass, we have also made claims about the absolute number of detections. Figures 2-5, 7-10, and Tables 2, 3 may under-estimate the number of possible planet detections for several reasons: contrast levels below $C_0=5\times10^{-7}$ may be possible for bright stars; the frequency of multiple planet systems may be greater than 28\%; systems with more than 4 Jovian planets may exist; our AO system model does not account for reduced differential chromatic aberrations when using a NIRWFS \citep{guyon_06}; instruments will also use the polarization signature of planets for detection \citep{roelfsema_10}; and, more than one planet formation mechanism seems likely. At the same time, we may have over-estimated the number of planet detections: a $dn_p/da \propto a^{-0.61}$ semimajor axis distribution may be too optimistic for wide-separation companions; it is doubtful that surveys will have sufficient time allocations to observe several hundred stars in each of the JHK filters; and, depending on the spectral-type, stars may require 2-3 hours of integration time each to reach contrast levels of order $C_0=10^{-7}$, therefore also limiting the number of targets observed. Further, the effects of orbit migration and planet-planet scattering can also alter the number of planet detections. \\

{\noindent}(2) We have applied a uniform multiplicity fraction to stars of all masses. Low-mass stars appear to have a binary fraction that is 10-20\% lower than solar-type stars as a consequence of natal gravitational interactions with cluster members \citep{fischer_marcy_92}. The binary fraction of high-mass stars is plausibly higher than 0.33, but this has yet to be measured \citep{raghavan_10}. The number of planet detections depends at the tens of percent level on the stellar mass-dependent binary fraction and whether the individual components of wide binaries are targeted. \\

{\noindent}(3) We have calculated the number of planet detections per star. The number of planet detections per time however is a function of target brightness. In practice, the effects of duty-cycle efficiency with respect to instrument calibration may further enhance the appeal of massive stars. \\

{\noindent}(4) Our simulations suggest that imaging planets that resemble the RV population will be unlikely around low-mass stars. We have considered a case where stars were chosen based on their age and brightness. However, another option to further improve the likelihood for discovery is to conduct a metallicity-biased survey. \citet{agol_07} discusses the prospects for such a strategy using space-based coronagraphs. Likewise, searches based on the existence of long-term RV trends may also be efficient \citep{crepp_johnson_11a}. \\

{\noindent}(5) We have applied number statistics for the age of FG stars to K-type and early M-type stars. The discovery of significantly more young, low-mass stars would improve the prospects for imaging planets in the immediate solar neighborhood. Observers are currently using activity-based youth indicators to identify more nearby low-mass targets \citep{liu_09,shkolnik_09}. \\


{\noindent}(6) We have extended the \citet{johnson_10_mass_met} results to stellar masses beyond $M_*=2.0M_{\odot}$, the range considered in their study. The planet occurrence rate may, however, saturate at high stellar mass. In this case, our calculations of the relative detectability for $\alpha=1.0$ and $\beta=1.2$ would over-estimate the relative promise of the most massive A-stars. However, given the number of super-Jupiters ($m_p\;\mbox{sin}\;i > 5 M_J$) discovered around massive K-giants with $M_* > 2.0M_{\odot}$, it does not appear that the relationship measured by \citet{johnson_10_mass_met} saturates in the range $2.0 < M_*/M_{\odot} < 3.0$ \citep{sato_10}. \\

{\noindent}(7) One challenge with measuring and interpreting planet detection rates is that dynamical effects, such as the gravitational impulse from nearby passing stars, can deplete wide separation planets at young ages \citep{veras_09}. Such encounters would preferentially effect low mass stars based on binding energy arguments \citep{weinberg_87}. In this case, detection rates around MK stars may be artificially low and mimic the results for an alternative formation scenario. 

\section{Summary and Concluding Remarks}
We have used Monte Carlo simulations to estimate the number of extrasolar planets that are directly detectable in the solar-neighborhood as a function of stellar mass with current and next-generation high-contrast imaging instruments. Starting with the most recent empirical findings for gas-giant planet occurrence-rates from Doppler RV surveys, we consider ground-based observations that survey volume-limited, and age-and-brightness selected samples of synthetic stars. Our calculations show that the probability for imaging a planet in the solar neighborhood lies heavily in favor of A-type stars as a result of strong correlations between star mass and planet properties. Such correlations are known to be present in the RV sample, and are required to reproduce direct imaging discoveries and non-discoveries to date (Fig. 2-5, 7). This result is independent of planet evolutionary model choice. It also holds when targets are pre-selected for youth and brightness, indicating that massive stars are ideal targets for reasons in addition to their natural youth and ability to serve as a bright beacon for AO correction. 

The same effects responsible for creating a multitude of detectable planets around massive stars conspire to reduce the number orbiting low-mass stars. 
Loosening target selection criterion to incorporate older and fainter stars, such as in the case of a volume-limited survey, can increase the number of planet detections around MK stars by a factor of several but at the expense of significantly lower efficiency. An AO system with a wave-front sensor operating in either of the I,Y,J bands can boost the number and rate of detections in this regime, suffering only a minor penalty with more massive stars. 

Using simulation input parameters that best reproduce observations in the solar neighborhood, we predict that surveys using instruments that generate contrast levels of $C_0\approx5\times10^{-6}$ will provide only a marginal increase to the known planet population at large separations, at most doubling the current number of detections. Surveys using instruments with ``extreme" AO that generate contrast levels of $C_0\approx5\times10^{-7}$ will provide a substantial gain, permitting statistical analyses with a sample of 12-81 new planet imaging discoveries (Fig. 8). 

To maximize the planet yield, we find it efficient to observe massive stars at distances beyond 50 pc, the canonical limit at which many observing campaigns are currently designed. Assuming non-ideal conditions, such as a bright sky background and decreasing planet semimajor axis distribution that truncates at 35 AU, dedicated observations can provide another factor of 2-3 in the number of directly imaged planets by extending survey selection to greater distances, for example detecting at least one companion for every 17 stars observed with mass, $2.0 < M_*/M_{\odot} \leq 2.5$, out to 87 pc with $C_0=5\times10^{-7}$ (Fig. 9). Searching closer to target stars enhances the number of detections in this regime by $\approx11\%$ when going from an IWA of $4 \;\lambda/D$ to $3 \; \lambda/D$. 

Stellar clusters represent a special case in high-contrast imaging because the constituent members have the same age and distance, neutralizing two important trade-offs between high-mass stars and low-mass stars. In this case, the number of detections is governed primarily by the stellar IMF, favoring low-mass stars by default (Fig. 10). If no correlations between star mass and planet properties exist, the detection rate becomes flat as a function of stellar mass, and may have a small negative slope because low-mass stars provide access to lower-mass planets. Simulations of a collection of stars resembling Tucana-Horologium show that this is the only case where M-stars can provide the most detections and also have the highest detection efficiencies. The inclusion of any correlations whatsoever between star mass and planet properties creates a positive slope, shifting the likelihood for detection back in favor of high-mass stars. 

At the same time, most nearby stellar clusters suffer from small number statistics, making it difficult to image a planet around its high-mass members, because the stellar IMF falls off faster than the planet occurrence rate grows. This effect is exacerbated by the fact that: half of cluster members are binaries and high-contrast programs usually avoid binaries; most stellar clusters are distant, $\approx$140 pc from the Sun, forcing the overall planet occurrence rate to take on its value at large separations; and the occurrence rate of ultra-wide separation companions appears to be very small, of order several percent. Thus, MK stars will continue to dominate the number of planet detections in distant stellar clusters containing $\lesssim$200 members irrespective of formation mechanism.\footnote{Notice however that aperture masking observations can access small star-planet separations and help to circumvent this bias by targeting the bright members of clusters \citep{kraus_11aas,hinkley_hr8799}.} Since observers have generally either conducted age-and-brightness selected surveys restricted to the solar neighborhood, or systematic observations of nearby clusters, the current direct imaging planet population forms a dichotomy between early and late-type stars (Table 4). We find that this dichotomy can be parsed according to the ratio of the effective spatial size of the young stellar group to the average distance of its members from the Sun. 

The planets orbiting HR 8799 and $\beta$ Pic have much smaller semimajor axes and masses than those found in stellar clusters, and more closely resemble the RV population. Further suggestive of similarity is the fact that planets have been found around these two A-stars, while there exists a dearth of detections around FGKM stars in the solar neighborhood despite dedicated efforts. The higher detection rate around A-stars compared to less massive stars is consistent with the correlations between star mass and planet properties found in the RV population. 

Since the detection of planets orbiting HR 8799 and $\beta$ Pictoris requires that strong correlations exist between star and planet properties, we propose that observers measure the slope of the detection rate curve for MK star cluster members. A flat or negative slope would indicate a separate formation channel, specifically one with a weak correspondence between star and planet properties, explaining the planets orbiting AB Pic, 1RSXJ1609, GSC 06214 and others. Such a diagnostic provides a complementary test to those involving measurement of metallicity correlations, semimajor axis distributions, and absolute occurrence rates.

Although planets will eventually be imaged around nearby FG stars, under no circumstances do we find that intermediate spectral-type hosts have the highest detection rates. The results are either skewed to one side or the other. Along with the aforementioned biases, this may help account for the current list of exoplanet imaging discoveries. Observations of A and M-stars can therefore provide the most leverage for discriminating between planet formation scenarios. Surveys are currently planned to specifically target stars in these mass regimes, using ``extreme" AO at Palomar \citep{hinkley_11_PASP} and deep AO observations at Keck (Bowler 2011, private communication). 

In closing, we note that the first handful of Doppler-detected planets showed a strong preference for formation around stars with an elevated metal content \citep{gonzalez_97}. Subsequent radial velocity observations have filled in this parameter space, quantifying the relationship between planet occurrence and host star metallicity using a statistically significant sample \citep{santos_03,fv_05,johnson_10_mass_met}. If the first few direct planet detections are a hint for which parameters govern formation at wide separations, an analogous situation in the field of high-contrast imaging may transpire involving correlations with star mass.

\vspace{0.1in}
We are grateful for discussions with Ruslan Belikov, whose questions during the Sagan symposium in Pasadena motivated us to perform this study. We thank Chas Beichman, Kelly Plummer, and Dennis Wittman for support using the Bluedot super-computing cluster at NExScI, and Eric Nielsen for helpful conversations about Monte Carlo simulations. Brendan Bowler, Adam Kraus, Lynne Hillenbrand, Sasha Hinkley, and Eric Nielsen read an early draft of this manuscript and provided valuable feedback that improved the presentation of our results. We thank the anonymous referee for additional helpful comments. This study made use of the NStED database. 

\begin{small}
\bibliographystyle{jtb}
\bibliography{ms.bib}

\begin{thebibliography}{}

\bibitem[\protect\astroncite{{Absil} and {Mawet}}{2010}]{absil_mawet_10}
{Absil}, O. and {Mawet}, D. (2010) ,
\newblock {\em \aapr} {\bf 18}, 317

\bibitem[\protect\astroncite{{Agol}}{2007}]{agol_07}
{Agol}, E. (2007) ,
\newblock {\em \mnras} {\bf 374}, 1271

\bibitem[\protect\astroncite{{Baraffe} et~al.}{2003}]{baraffe_03}
{Baraffe}, I., {Chabrier}, G., {Barman}, T.~S., {Allard}, F., and {Hauschildt},
  P.~H. (2003) ,
\newblock {\em \aap} {\bf 402}, 701

\bibitem[\protect\astroncite{{Baranec}}{2008}]{baranec_08}
{Baranec}, C. (2008) ,
\newblock In {\em Society of Photo-Optical Instrumentation Engineers (SPIE)
  Conference Series}, Vol. 7015 of {\em Presented at the Society of
  Photo-Optical Instrumentation Engineers (SPIE) Conference}

\bibitem[\protect\astroncite{{Barrado y Navascues}}{1998}]{barrado_98}
{Barrado y Navascues}, D. (1998) ,
\newblock {\em \aap} {\bf 339}, 831

\bibitem[\protect\astroncite{{Beichman} et~al.}{2010}]{beichman_10}
{Beichman}, C.~A., {Krist}, J., {Trauger}, J.~T., {Greene}, T., {Oppenheimer},
  B., {Sivaramakrishnan}, A., {Doyon}, R., {Boccaletti}, A., {Barman}, T.~S.,
  and {Rieke}, M. (2010) ,
\newblock {\em \pasp} {\bf 122}, 162

\bibitem[\protect\astroncite{{Biller} et~al.}{2007}]{biller_07}
{Biller}, B.~A., {Close}, L.~M., {Masciadri}, E., {Nielsen}, E., {Lenzen}, R.,
  {Brandner}, W., {McCarthy}, D., {Hartung}, M., {Kellner}, S., {Mamajek}, E.,
  {Henning}, T., {Miller}, D., {Kenworthy}, M., and {Kulesa}, C. (2007) ,
\newblock {\em \apjs} {\bf 173}, 143

\bibitem[\protect\astroncite{{Biller} et~al.}{2010}]{biller_10}
{Biller}, B.~A., {Wahhaj}, Z., {Liu}, M., {Chun}, M., {Close}, L., {Ftaclas},
  C., {Hartung}, M., {Hayward}, T., {Nielsen}, E., {Toomey}, D., and {NICI
  Planet-Finding Campaign team} (2010) ,
\newblock In {\em Bulletin of the American Astronomical Society}, Vol.~41 of
  {\em Bulletin of the American Astronomical Society}, pp. 457--+

\bibitem[\protect\astroncite{{Boley}}{2009}]{boley_09}
{Boley}, A.~C. (2009) ,
\newblock {\em \apjl} {\bf 695}, L53

\bibitem[\protect\astroncite{{Boss}}{1997}]{boss_97}
{Boss}, A.~P. (1997) ,
\newblock {\em Science} {\bf 276}, 1836

\bibitem[\protect\astroncite{{Boss}}{2011}]{boss_11}
{Boss}, A.~P. (2011) ,
\newblock {\em ArXiv e-prints}

\bibitem[\protect\astroncite{{Bouchez} et~al.}{2009}]{bouchez_09}
{Bouchez}, A.~H., {Dekany}, R.~G., {Angione}, J.~R., {Baranec}, C., {Bui}, K.,
  {Burruss}, R.~S., {Crepp}, J.~R., {Croner}, E.~E., {Cromer}, J.~L.,
  {Guiwits}, S.~R., {Hale}, D.~D.~S., {Henning}, J.~R., {Palmer}, D.,
  {Roberts}, J.~E., {Troy}, M., {Truong}, T.~N., and {Zolkower}, J. (2009) ,
\newblock In {\em Society of Photo-Optical Instrumentation Engineers (SPIE)
  Conference Series}, Vol. 7439 of {\em Presented at the Society of
  Photo-Optical Instrumentation Engineers (SPIE) Conference}

\bibitem[\protect\astroncite{{Bowler} et~al.}{2010a}]{bowler_10}
{Bowler}, B.~P., {Johnson}, J.~A., {Marcy}, G.~W., {Henry}, G.~W., {Peek},
  K.~M.~G., {Fischer}, D.~A., {Clubb}, K.~I., {Liu}, M.~C., {Reffert}, S.,
  {Schwab}, C., and {Lowe}, T.~B. (2010a) ,
\newblock {\em \apj} {\bf 709}, 396

\bibitem[\protect\astroncite{{Bowler} et~al.}{2010b}]{bowler_10_hr8799b}
{Bowler}, B.~P., {Liu}, M.~C., {Dupuy}, T.~J., and {Cushing}, M.~C. (2010b) ,
\newblock {\em \apj} {\bf 723}, 850

\bibitem[\protect\astroncite{{Carson} et~al.}{2006}]{carson_06}
{Carson}, J.~C., {Eikenberry}, S.~S., {Smith}, J.~J., and {Cordes}, J.~M.
  (2006) ,
\newblock {\em \aj} {\bf 132}, 1146

\bibitem[\protect\astroncite{{Chauvin} et~al.}{2005}]{chauvin_05_abpic}
{Chauvin}, G., {Lagrange}, A.-M., {Zuckerman}, B., {Dumas}, C., {Mouillet}, D.,
  {Song}, I., {Beuzit}, J.-L., {Lowrance}, P., and {Bessell}, M.~S. (2005) ,
\newblock {\em \aap} {\bf 438}, L29

\bibitem[\protect\astroncite{{Crepp} et~al.}{2010}]{crepp_10}
{Crepp}, J., {Serabyn}, E., {Carson}, J., {Ge}, J., and {Kravchenko}, I. (2010)
  ,
\newblock {\em \apj} {\bf 715}, 1533

\bibitem[\protect\astroncite{{Crepp} and {Johnson}}{2011}]{crepp_johnson_11a}
{Crepp}, J.~R. and {Johnson}, J.~A. (2011) ,
\newblock In {\em American Astronomical Society Meeting Abstracts}, Vol. 217 of
  {\em American Astronomical Society Meeting Abstracts}, p. 302.05

\bibitem[\protect\astroncite{{Crepp} et~al.}{2009}]{crepp_09}
{Crepp}, J.~R., {Mahadevan}, S., and {Ge}, J. (2009) ,
\newblock {\em \apj} {\bf 702}, 672

\bibitem[\protect\astroncite{{Crepp} et~al.}{2011}]{crepp_11}
{Crepp}, J.~R., {Pueyo}, L., {Brenner}, D., {Oppenheimer}, B.~R., {Zimmerman},
  N., {Hinkley}, S., {Parry}, I., {King}, D., {Vasisht}, G., {Beichman}, C.,
  {Hillenbrand}, L., {Dekany}, R., {Shao}, M., {Burruss}, R., {Roberts}, L.~C.,
  {Bouchez}, A., {Roberts}, J., and {Soummer}, R. (2011) ,
\newblock {\em \apj} {\bf 729}, 132

\bibitem[\protect\astroncite{{Crida} et~al.}{2009}]{crida_09}
{Crida}, A., {Masset}, F., and {Morbidelli}, A. (2009) ,
\newblock {\em \apjl} {\bf 705}, L148

\bibitem[\protect\astroncite{{Cumming} et~al.}{2008}]{cumming_08}
{Cumming}, A., {Butler}, R.~P., {Marcy}, G.~W., {Vogt}, S.~S., {Wright}, J.~T.,
  and {Fischer}, D.~A. (2008) ,
\newblock {\em \pasp} {\bf 120}, 531

\bibitem[\protect\astroncite{{Currie} et~al.}{2011}]{currie_11}
{Currie}, T., {Burrows}, A.~S., {Itoh}, Y., {Matsumura}, S., {Fukagawa}, M.,
  {Apai}, D., {Madhusudhan}, N., {Hinz}, P.~M., {Rodigas}, T., {Kasper}, M.,
  {Pyo}, T., and {Ogino}, S. (2011) ,
\newblock {\em ArXiv e-prints}

\bibitem[\protect\astroncite{{da Silva} et~al.}{2006}]{dasilva_06}
{da Silva}, R., {Udry}, S., {Bouchy}, F., {Mayor}, M., {Moutou}, C., {Pont},
  F., {Queloz}, D., {Santos}, N.~C., {S{\'e}gransan}, D., and {Zucker}, S.
  (2006) ,
\newblock {\em \aap} {\bf 446}, 717

\bibitem[\protect\astroncite{{Dodson-Robinson} et~al.}{2009}]{dodson_09}
{Dodson-Robinson}, S.~E., {Veras}, D., {Ford}, E.~B., and {Beichman}, C.~A.
  (2009) ,
\newblock {\em \apj} {\bf 707}, 79

\bibitem[\protect\astroncite{{Dohlen} et~al.}{2006}]{dohlen_06}
{Dohlen}, K., {Beuzit}, J.-L., {Feldt}, M., {Mouillet}, D., {Puget}, P.,
  {Antichi}, J., {Baruffolo}, A., {Baudoz}, P., {Berton}, A., {Boccaletti}, A.,
  {Carbillet}, M., {Charton}, J., {Claudi}, R., {Downing}, M., {Fabron}, C.,
  {Feautrier}, P., {Fedrigo}, E., {Fusco}, T., {Gach}, J.-L., {Gratton}, R.,
  {Hubin}, N., {Kasper}, M., {Langlois}, M., {Longmore}, A., {Moutou}, C.,
  {Petit}, C., {Pragt}, J., {Rabou}, P., {Rousset}, G., {Saisse}, M., {Schmid},
  H.-M., {Stadler}, E., {Stamm}, D., {Turatto}, M., {Waters}, R., and {Wildi},
  F. (2006) ,
\newblock In {\em Ground-based and Airborne Instrumentation for Astronomy.
  Edited by McLean, Ian S.; Iye, Masanori. Proceedings of the SPIE, Volume
  6269, pp. 62690Q (2006).}, Vol. 6269 of {\em Presented at the Society of
  Photo-Optical Instrumentation Engineers (SPIE) Conference}

\bibitem[\protect\astroncite{{Durisen} et~al.}{2007}]{durisen_07}
{Durisen}, R.~H., {Hartquist}, T.~W., and {Pickett}, M.~K. (2007) ,
\newblock {\em ArXiv e-prints} 709

\bibitem[\protect\astroncite{{Ehrenreich} et~al.}{2010}]{ehrenreich_10}
{Ehrenreich}, D., {Lagrange}, A., {Montagnier}, G., {Chauvin}, G., {Galland},
  F., {Beuzit}, J., and {Rameau}, J. (2010) ,
\newblock {\em ArXiv e-prints}

\bibitem[\protect\astroncite{{Fischer} et~al.}{2005}]{fischer_05}
{Fischer}, D.~A., {Laughlin}, G., {Butler}, P., {Marcy}, G., {Johnson}, J.,
  {Henry}, G., {Valenti}, J., {Vogt}, S., {Ammons}, M., {Robinson}, S.,
  {Spear}, G., {Strader}, J., {Driscoll}, P., {Fuller}, A., {Johnson}, T.,
  {Manrao}, E., {McCarthy}, C., {Mu{\~n}oz}, M., {Tah}, K.~L., {Wright}, J.,
  {Ida}, S., {Sato}, B., {Toyota}, E., and {Minniti}, D. (2005) ,
\newblock {\em \apj} {\bf 620}, 481

\bibitem[\protect\astroncite{{Fischer} and {Marcy}}{1992}]{fischer_marcy_92}
{Fischer}, D.~A. and {Marcy}, G.~W. (1992) ,
\newblock {\em \apj} {\bf 396}, 178

\bibitem[\protect\astroncite{{Fischer} et~al.}{2008}]{fischer_08}
{Fischer}, D.~A., {Marcy}, G.~W., {Butler}, R.~P., {Vogt}, S.~S., {Laughlin},
  G., {Henry}, G.~W., {Abouav}, D., {Peek}, K.~M.~G., {Wright}, J.~T.,
  {Johnson}, J.~A., {McCarthy}, C., and {Isaacson}, H. (2008) ,
\newblock {\em \apj} {\bf 675}, 790

\bibitem[\protect\astroncite{{Fischer} and {Valenti}}{2005}]{fv_05}
{Fischer}, D.~A. and {Valenti}, J. (2005) ,
\newblock {\em \apj} {\bf 622}, 1102

\bibitem[\protect\astroncite{{Fortney} et~al.}{2008}]{fortney_08}
{Fortney}, J.~J., {Marley}, M.~S., {Saumon}, D., and {Lodders}, K. (2008) ,
\newblock {\em \apj} {\bf 683}, 1104

\bibitem[\protect\astroncite{{Girardi} et~al.}{2002}]{girardi_02}
{Girardi}, L., {Bertelli}, G., {Bressan}, A., {Chiosi}, C., {Groenewegen},
  M.~A.~T., {Marigo}, P., {Salasnich}, B., and {Weiss}, A. (2002) ,
\newblock {\em \aap} {\bf 391}, 195

\bibitem[\protect\astroncite{{Give'on} et~al.}{2007}]{giveon_07}
{Give'on}, A., {Belikov}, R., {Shaklan}, S., and {Kasdin}, J. (2007) ,
\newblock {\em Optics Express} {\bf 15}, 12338

\bibitem[\protect\astroncite{{Gonzalez}}{1997}]{gonzalez_97}
{Gonzalez}, G. (1997) ,
\newblock {\em \mnras} {\bf 285}, 403

\bibitem[\protect\astroncite{{Gray} et~al.}{2003}]{gray_03}
{Gray}, R.~O., {Corbally}, C.~J., {Garrison}, R.~F., {McFadden}, M.~T., and
  {Robinson}, P.~E. (2003) ,
\newblock {\em \aj} {\bf 126}, 2048

\bibitem[\protect\astroncite{{Guyon} et~al.}{2006}]{guyon_06}
{Guyon}, O., {Pluzhnik}, E.~A., {Kuchner}, M.~J., {Collins}, B., and {Ridgway},
  S.~T. (2006) ,
\newblock {\em \apjs} {\bf 167}, 81

\bibitem[\protect\astroncite{{Hinkley} et~al.}{2011a}]{hinkley_hr8799}
{Hinkley}, S., {Carpenter}, J.~M., {Ireland}, M.~J., and {Kraus}, A.~L. (2011a)
  ,
\newblock {\em ArXiv e-prints}

\bibitem[\protect\astroncite{{Hinkley} et~al.}{2011b}]{hinkley_11_PASP}
{Hinkley}, S., {Oppenheimer}, B.~R., {Zimmerman}, N., {Brenner}, D., {Parry},
  I.~R., {Crepp}, J.~R., {Vasisht}, G., {Ligon}, E., {King}, D., {Soummer}, R.,
  {Sivaramakrishnan}, A., {Beichman}, C., {Shao}, M., {Roberts}, L.~C.,
  {Bouchez}, A., {Dekany}, R., {Pueyo}, L., {Roberts}, J.~E., {Lockhart}, T.,
  {Zhai}, C., {Shelton}, C., and {Burruss}, R. (2011b) ,
\newblock {\em \pasp} {\bf 123}, 74

\bibitem[\protect\astroncite{{Holman} and {Wiegert}}{1999}]{holman_wiegert_99}
{Holman}, M.~J. and {Wiegert}, P.~A. (1999) ,
\newblock {\em \aj} {\bf 117}, 621

\bibitem[\protect\astroncite{{Ida} and {Lin}}{2004}]{ida_lin_04}
{Ida}, S. and {Lin}, D.~N.~C. (2004) ,
\newblock {\em \apj} {\bf 616}, 567

\bibitem[\protect\astroncite{{Ireland} et~al.}{2011}]{ireland_11}
{Ireland}, M.~J., {Kraus}, A., {Martinache}, F., {Law}, N., and {Hillenbrand},
  L.~A. (2011) ,
\newblock {\em \apj} {\bf 726}, 113

\bibitem[\protect\astroncite{{Janson} et~al.}{2011}]{janson_11}
{Janson}, M., {Carson}, J., {Thalmann}, C., {McElwain}, M.~W., {Goto}, M.,
  {Crepp}, J., {Wisniewski}, J., {Abe}, L., {Brandner}, W., {Burrows}, A.,
  {Egner}, S., {Feldt}, M., {Grady}, C.~A., {Golota}, T., {Guyon}, O.,
  {Hashimoto}, J., {Hayano}, Y., {Hayashi}, M., {Hayashi}, S., {Henning}, T.,
  {Hodapp}, K.~W., {Ishii}, M., {Iye}, M., {Kandori}, R., {Knapp}, G.~R.,
  {Kudo}, T., {Kusakabe}, N., {Kuzuhara}, M., {Matsuo}, T., {Mayama}, S.,
  {Miyama}, S., {Morino}, J., {Moro-Mart{\'{\i}}n}, A., {Nishimura}, T., {Pyo},
  T., {Serabyn}, E., {Suto}, H., {Suzuki}, R., {Takami}, M., {Takato}, N.,
  {Terada}, H., {Tofflemire}, B., {Tomono}, D., {Turner}, E.~L., {Watanabe},
  M., {Yamada}, T., {Takami}, H., {Usuda}, T., and {Tamura}, M. (2011) ,
\newblock {\em \apj} {\bf 728}, 85

\bibitem[\protect\astroncite{{Johnson}}{2009}]{johnson_09_review}
{Johnson}, J.~A. (2009) ,
\newblock {\em \pasp} {\bf 121}, 309

\bibitem[\protect\astroncite{{Johnson} et~al.}{2010a}]{johnson_10_mass_met}
{Johnson}, J.~A., {Aller}, K.~M., {Howard}, A.~W., and {Crepp}, J.~R. (2010a) ,
\newblock {\em \pasp} {\bf 122}, 905

\bibitem[\protect\astroncite{{Johnson} et~al.}{2010b}]{johnson_10_subgiant}
{Johnson}, J.~A., {Bowler}, B.~P., {Howard}, A.~W., {Henry}, G.~W., {Marcy},
  G.~W., {Isaacson}, H., {Brewer}, J.~M., {Fischer}, D.~A., {Morton}, T.~D.,
  and {Crepp}, J.~R. (2010b) ,
\newblock {\em ArXiv e-prints}

\bibitem[\protect\astroncite{{Johnson} et~al.}{2008}]{johnson_08}
{Johnson}, J.~A., {Marcy}, G.~W., {Fischer}, D.~A., {Wright}, J.~T., {Reffert},
  S., {Kregenow}, J.~M., {Williams}, P.~K.~G., and {Peek}, K.~M.~G. (2008) ,
\newblock {\em \apj} {\bf 675}, 784

\bibitem[\protect\astroncite{{Jorissen} et~al.}{2001}]{jorissen_01}
{Jorissen}, A., {Mayor}, M., and {Udry}, S. (2001) ,
\newblock {\em \aap} {\bf 379}, 992

\bibitem[\protect\astroncite{{Kalas} et~al.}{2008}]{kalas_08}
{Kalas}, P., {Graham}, J.~R., {Chiang}, E., {Fitzgerald}, M.~P., {Clampin}, M.,
  {Kite}, E.~S., {Stapelfeldt}, K., {Marois}, C., and {Krist}, J. (2008) ,
\newblock {\em Science} {\bf 322}, 1345

\bibitem[\protect\astroncite{{Kataria} and {Simon}}{2010}]{kataria_10}
{Kataria}, T. and {Simon}, M. (2010) ,
\newblock {\em \aj} {\bf 140}, 206

\bibitem[\protect\astroncite{{Kennedy} and {Kenyon}}{2008}]{kennedy_08}
{Kennedy}, G.~M. and {Kenyon}, S.~J. (2008) ,
\newblock {\em \apj} {\bf 673}, 502

\bibitem[\protect\astroncite{{Kratter} et~al.}{2010}]{kratter_10}
{Kratter}, K.~M., {Murray-Clay}, R.~A., and {Youdin}, A.~N. (2010) ,
\newblock {\em \apj} {\bf 710}, 1375

\bibitem[\protect\astroncite{{Kraus} and {Ireland}}{2011}]{kraus_11aas}
{Kraus}, A.~L. and {Ireland}, M.~J. (2011) ,
\newblock In {\em American Astronomical Society Meeting Abstracts 217}, Vol.~43
  of {\em Bulletin of the American Astronomical Society}, p. 302.03

\bibitem[\protect\astroncite{{Lafreni{\`e}re} et~al.}{2007}]{lafreniere_07}
{Lafreni{\`e}re}, D., {Doyon}, R., {Marois}, C., {Nadeau}, D., {Oppenheimer},
  B.~R., {Roche}, P.~F., {Rigaut}, F., {Graham}, J.~R., {Jayawardhana}, R.,
  {Johnstone}, D., {Kalas}, P.~G., {Macintosh}, B., and {Racine}, R. (2007) ,
\newblock {\em \apj} {\bf 670}, 1367

\bibitem[\protect\astroncite{{Lafreni{\`e}re} et~al.}{2011}]{lafreniere_11}
{Lafreni{\`e}re}, D., {Jayawardhana}, R., {Janson}, M., {Helling}, C., {Witte},
  S., and {Hauschildt}, P. (2011) ,
\newblock {\em ArXiv e-prints}

\bibitem[\protect\astroncite{{Lagrange} et~al.}{2010}]{lagrange_10}
{Lagrange}, A., {Bonnefoy}, M., {Chauvin}, G., {Apai}, D., {Ehrenreich}, D.,
  {Boccaletti}, A., {Gratadour}, D., {Rouan}, D., {Mouillet}, D., {Lacour}, S.,
  and {Kasper}, M. (2010) ,
\newblock {\em ArXiv e-prints}

\bibitem[\protect\astroncite{{Latyshev}}{1978}]{latyshev_78}
{Latyshev}, I.~N. (1978) ,
\newblock {\em Soviet Astronomy} {\bf 22}, 186

\bibitem[\protect\astroncite{{Laughlin}}{2000}]{laughlin_00}
{Laughlin}, G. (2000) ,
\newblock {\em \apj} {\bf 545}, 1064

\bibitem[\protect\astroncite{{Leconte} et~al.}{2010}]{leconte_10}
{Leconte}, J., {Soummer}, R., {Hinkley}, S., {Oppenheimer}, B.~R.,
  {Sivaramakrishnan}, A., {Brenner}, D., {Kuhn}, J., {Lloyd}, J.~P., {Perrin},
  M.~D., {Makidon}, R., {Roberts}, Jr., L.~C., {Graham}, J.~R., {Simon}, M.,
  {Brown}, R.~A., {Zimmerman}, N., {Chabrier}, G., and {Baraffe}, I. (2010) ,
\newblock {\em \apj} {\bf 716}, 1551

\bibitem[\protect\astroncite{{Liu} et~al.}{2009}]{liu_09}
{Liu}, M.~C., {Wahhaj}, Z., {Biller}, B., {Shkolnik}, E., {Chun}, M.,
  {Ftaclas}, C., {Toomey}, D., {Close}, L., {Nielsen}, E., {Hayward}, T.,
  {Hartung}, M., and {Artigau}, E. (2009) ,
\newblock In {\em American Institute of Physics Conference Series}.
  ({E.~Stempels} ed.), Vol. 1094 of {\em American Institute of Physics
  Conference Series}, pp. 461--464

\bibitem[\protect\astroncite{{Liu} et~al.}{2010}]{liu_10_spie}
{Liu}, M.~C., {Wahhaj}, Z., {Biller}, B.~A., {Nielsen}, E.~L., {Chun}, M.,
  {Close}, L.~M., {Ftaclas}, C., {Hartung}, M., {Hayward}, T.~L., {Clarke}, F.,
  {Reid}, I.~N., {Shkolnik}, E.~L., {Tecza}, M., {Thatte}, N., {Alencar}, S.,
  {Artymowicz}, P., {Boss}, A., {Burrows}, A., {de Gouveia Dal Pino}, E.,
  {Gregorio-Hetem}, J., {Ida}, S., {Kuchner}, M.~J., {Lin}, D., and {Toomey},
  D. (2010) ,
\newblock In {\em Society of Photo-Optical Instrumentation Engineers (SPIE)
  Conference Series}, Vol. 7736 of {\em Presented at the Society of
  Photo-Optical Instrumentation Engineers (SPIE) Conference}

\bibitem[\protect\astroncite{{Lovis} and {Mayor}}{2007}]{lovis_07}
{Lovis}, C. and {Mayor}, M. (2007) ,
\newblock {\em \aap} {\bf 472}, 657

\bibitem[\protect\astroncite{{Macintosh} et~al.}{2006}]{macintosh_GPI_06}
{Macintosh}, B., {Graham}, J., {Palmer}, D., {Doyon}, R., {Gavel}, D.,
  {Larkin}, J., {Oppenheimer}, B., {Saddlemyer}, L., {Wallace}, J.~K.,
  {Bauman}, B., {Evans}, J., {Erikson}, D., {Morzinski}, K., {Phillion}, D.,
  {Poyneer}, L., {Sivaramakrishnan}, A., {Soummer}, R., {Thibault}, S., and
  {Veran}, J.-P. (2006) ,
\newblock In {\em Advances in Adaptive Optics II. Edited by Ellerbroek, Brent
  L.; Bonaccini Calia, Domenico. Proceedings of the SPIE, Volume 6272, pp.
  62720L (2006).}, Vol. 6272 of {\em Presented at the Society of Photo-Optical
  Instrumentation Engineers (SPIE) Conference}

\bibitem[\protect\astroncite{{Madhusudhan} et~al.}{2011}]{madhusudhan_11}
{Madhusudhan}, N., {Burrows}, A., and {Currie}, T. (2011) ,
\newblock {\em ArXiv e-prints}

\bibitem[\protect\astroncite{{Makarov}}{2007a}]{makarov_07}
{Makarov}, V.~V. (2007a) ,
\newblock {\em \apj} {\bf 658}, 480

\bibitem[\protect\astroncite{{Makarov}}{2007b}]{makarov_07b}
{Makarov}, V.~V. (2007b) ,
\newblock {\em \apjs} {\bf 169}, 105

\bibitem[\protect\astroncite{{Malbet} et~al.}{1995}]{malbet_95}
{Malbet}, F., {Yu}, J.~W., and {Shao}, M. (1995) ,
\newblock {\em \pasp} {\bf 107}, 386

\bibitem[\protect\astroncite{{Mamajek} and
  {Hillenbrand}}{2008}]{mamajek_hillenbrand_08}
{Mamajek}, E.~E. and {Hillenbrand}, L.~A. (2008) ,
\newblock {\em \apj} {\bf 687}, 1264

\bibitem[\protect\astroncite{{Marcy} et~al.}{1999}]{marcy_99}
{Marcy}, G.~W., {Butler}, R.~P., {Vogt}, S.~S., {Fischer}, D., and {Liu}, M.~C.
  (1999) ,
\newblock {\em \apj} {\bf 520}, 239

\bibitem[\protect\astroncite{{Marcy} et~al.}{2008}]{marcy_08}
{Marcy}, G.~W., {Butler}, R.~P., {Vogt}, S.~S., {Fischer}, D.~A., {Wright},
  J.~T., {Johnson}, J.~A., {Tinney}, C.~G., {Jones}, H.~R.~A., {Carter}, B.~D.,
  {Bailey}, J., {O'Toole}, S.~J., and {Upadhyay}, S. (2008) ,
\newblock {\em Physica Scripta Volume T} {\bf 130(1)}, 014001

\bibitem[\protect\astroncite{{Marley} et~al.}{2007}]{marley_07}
{Marley}, M.~S., {Fortney}, J.~J., {Hubickyj}, O., {Bodenheimer}, P., and
  {Lissauer}, J.~J. (2007) ,
\newblock {\em \apj} {\bf 655}, 541

\bibitem[\protect\astroncite{{Marois} et~al.}{2006}]{marois_06}
{Marois}, C., {Lafreni{\`e}re}, D., {Doyon}, R., {Macintosh}, B., and {Nadeau},
  D. (2006) ,
\newblock {\em \apj} {\bf 641}, 556

\bibitem[\protect\astroncite{{Marois} et~al.}{2008}]{marois_08}
{Marois}, C., {Macintosh}, B., {Barman}, T., {Zuckerman}, B., {Song}, I.,
  {Patience}, J., {Lafreni{\`e}re}, D., and {Doyon}, R. (2008) ,
\newblock {\em Science} {\bf 322}, 1348

\bibitem[\protect\astroncite{{Marois} et~al.}{2010}]{marois_10}
{Marois}, C., {Zuckerman}, B., {Konopacky}, Q.~M., {Macintosh}, B., and
  {Barman}, T. (2010) ,
\newblock {\em \nat} {\bf 468}, 1080

\bibitem[\protect\astroncite{{Masciadri} et~al.}{2005}]{masciadri_05}
{Masciadri}, E., {Mundt}, R., {Henning}, T., {Alvarez}, C., and {Barrado y
  Navascu{\'e}s}, D. (2005) ,
\newblock {\em \apj} {\bf 625}, 1004

\bibitem[\protect\astroncite{{Mayor} et~al.}{2009}]{mayor_09}
{Mayor}, M., {Udry}, S., {Lovis}, C., {Pepe}, F., {Queloz}, D., {Benz}, W.,
  {Bertaux}, J., {Bouchy}, F., {Mordasini}, C., and {Segransan}, D. (2009) ,
\newblock {\em \aap} {\bf 493}, 639

\bibitem[\protect\astroncite{{Metchev} and
  {Hillenbrand}}{2009}]{metchev_hillenbrand_09}
{Metchev}, S.~A. and {Hillenbrand}, L.~A. (2009) ,
\newblock {\em \apjs} {\bf 181}, 62

\bibitem[\protect\astroncite{{Miller} and {Scalo}}{1979}]{miller_scalo_79}
{Miller}, G.~E. and {Scalo}, J.~M. (1979) ,
\newblock {\em \apjs} {\bf 41}, 513

\bibitem[\protect\astroncite{{Murray} and
  {Dermott}}{2000}]{murray_dermott_book}
{Murray}, C.~D. and {Dermott}, S.~F. (2000) ,
\newblock {\em {Solar System Dynamics}}

\bibitem[\protect\astroncite{{Nielsen} and {Close}}{2010}]{nielsen_10}
{Nielsen}, E.~L. and {Close}, L.~M. (2010) ,
\newblock {\em \apj} {\bf 717}, 878

\bibitem[\protect\astroncite{{Nielsen} et~al.}{2008}]{nielsen_08}
{Nielsen}, E.~L., {Close}, L.~M., {Biller}, B.~A., {Masciadri}, E., and
  {Lenzen}, R. (2008) ,
\newblock {\em \apj} {\bf 674}, 466

\bibitem[\protect\astroncite{{Nordstr{\"o}m} et~al.}{2004}]{nordstrom_04}
{Nordstr{\"o}m}, B., {Mayor}, M., {Andersen}, J., {Holmberg}, J., {Pont}, F.,
  {J{\o}rgensen}, B.~R., {Olsen}, E.~H., {Udry}, S., and {Mowlavi}, N. (2004) ,
\newblock {\em \aap} {\bf 418}, 989

\bibitem[\protect\astroncite{{Noyes} et~al.}{1984}]{noyes_84}
{Noyes}, R.~W., {Hartmann}, L.~W., {Baliunas}, S.~L., {Duncan}, D.~K., and
  {Vaughan}, A.~H. (1984) ,
\newblock {\em \apj} {\bf 279}, 763

\bibitem[\protect\astroncite{{Oppenheimer} and
  {Hinkley}}{2009}]{oppenheimer_hinkley_09}
{Oppenheimer}, B.~R. and {Hinkley}, S. (2009) ,
\newblock {\em \araa} {\bf 47}, 253

\bibitem[\protect\astroncite{{Pollack} et~al.}{1996}]{pollack_et_al_96}
{Pollack}, J.~B., {Hubickyj}, O., {Bodenheimer}, P., {Lissauer}, J.~J.,
  {Podolak}, M., and {Greenzweig}, Y. (1996) ,
\newblock {\em Icarus} {\bf 124}, 62

\bibitem[\protect\astroncite{{Raghavan} et~al.}{2010}]{raghavan_10}
{Raghavan}, D., {McAlister}, H.~A., {Henry}, T.~J., {Latham}, D.~W., {Marcy},
  G.~W., {Mason}, B.~D., {Gies}, D.~R., {White}, R.~J., and {ten Brummelaar},
  T.~A. (2010) ,
\newblock {\em ArXiv e-prints}

\bibitem[\protect\astroncite{{Reid} et~al.}{2002}]{reid_02}
{Reid}, I.~N., {Gizis}, J.~E., and {Hawley}, S.~L. (2002) ,
\newblock {\em \aj} {\bf 124}, 2721

\bibitem[\protect\astroncite{{Reid} et~al.}{2007}]{reid_07}
{Reid}, I.~N., {Turner}, E.~L., {Turnbull}, M.~C., {Mountain}, M., and
  {Valenti}, J.~A. (2007) ,
\newblock {\em \apj} {\bf 665}, 767

\bibitem[\protect\astroncite{{Rigaut} et~al.}{1992}]{rigaut_92}
{Rigaut}, F., {Cuby}, J.~G., {Caes}, M., {Monin}, J.~L., {Vittot}, M.,
  {Richard}, J.~C., {Rousset}, G., and {Lena}, P. (1992) ,
\newblock {\em \aap} {\bf 259}, L57

\bibitem[\protect\astroncite{{Roelfsema} et~al.}{2010}]{roelfsema_10}
{Roelfsema}, R., {Schmid}, H.~M., {Pragt}, J., {Gisler}, D., {Waters}, R.,
  {Bazzon}, A., {Baruffolo}, A., {Beuzit}, J., {Boccaletti}, A., {Charton}, J.,
  {Cumani}, C., {Dohlen}, K., {Downing}, M., {Elswijk}, E., {Feldt}, M.,
  {Groothuis}, C., {de Haan}, M., {Hanenburg}, H., {Hubin}, N., {Joos}, F.,
  {Kasper}, M., {Keller}, C., {Kragt}, J., {Lizon}, J., {Mouillet}, D.,
  {Pavlov}, A., {Rigal}, F., {Rochat}, S., {Salasnich}, B., {Steiner}, P.,
  {Thalmann}, C., {Venema}, L., and {Wildi}, F. (2010) ,
\newblock In {\em Society of Photo-Optical Instrumentation Engineers (SPIE)
  Conference Series}, Vol. 7735 of {\em Presented at the Society of
  Photo-Optical Instrumentation Engineers (SPIE) Conference}

\bibitem[\protect\astroncite{{Santos} et~al.}{2003}]{santos_03}
{Santos}, N.~C., {Israelian}, G., {Mayor}, M., {Rebolo}, R., and {Udry}, S.
  (2003) ,
\newblock {\em \aap} {\bf 398}, 363

\bibitem[\protect\astroncite{{Santos} et~al.}{2010}]{santos_10}
{Santos}, N.~C., {Mayor}, M., {Benz}, W., {Bouchy}, F., {Figueira}, P., {Lo
  Curto}, G., {Lovis}, C., {Melo}, C., {Moutou}, C., {Naef}, D., {Pepe}, F.,
  {Queloz}, D., {Sousa}, S.~G., and {Udry}, S. (2010) ,
\newblock {\em \aap} {\bf 512}, A47+

\bibitem[\protect\astroncite{{Sato} et~al.}{2010}]{sato_10}
{Sato}, B., {Omiya}, M., {Liu}, Y., {Harakawa}, H., {Izumiura}, H., {Kambe},
  E., {Toyota}, E., {Murata}, D., {Lee}, B., {Masuda}, S., {Takeda}, Y.,
  {Yoshida}, M., {Itoh}, Y., {Ando}, H., {Kokubo}, E., {Ida}, S., {Zhao}, G.,
  and {Han}, I. (2010) ,
\newblock {\em ArXiv e-prints}

\bibitem[\protect\astroncite{{Scharf} and {Menou}}{2009}]{scharf_09}
{Scharf}, C. and {Menou}, K. (2009) ,
\newblock {\em \apjl} {\bf 693}, L113

\bibitem[\protect\astroncite{{Seager} and {Deming}}{2010}]{seager_10}
{Seager}, S. and {Deming}, D. (2010) ,
\newblock {\em \araa} {\bf 48}, 631

\bibitem[\protect\astroncite{{Shkolnik} et~al.}{2009}]{shkolnik_09}
{Shkolnik}, E., {Liu}, M.~C., and {Reid}, I.~N. (2009) ,
\newblock {\em \apj} {\bf 699}, 649

\bibitem[\protect\astroncite{{Soderblom}}{2010}]{soderblom_10}
{Soderblom}, D.~R. (2010) ,
\newblock {\em \araa} {\bf 48}, 581

\bibitem[\protect\astroncite{{Tamura}}{2009}]{tamura_09}
{Tamura}, M. (2009) ,
\newblock In {\em American Institute of Physics Conference Series}.
  ({T.~Usuda, M.~Tamura, \& M.~Ishii} ed.), Vol. 1158 of {\em American
  Institute of Physics Conference Series}, pp. 11--16

\bibitem[\protect\astroncite{{Thalmann} et~al.}{2009}]{thalmann_09}
{Thalmann}, C., {Carson}, J., {Janson}, M., {Goto}, M., {McElwain}, M.,
  {Egner}, S., {Feldt}, M., {Hashimoto}, J., {Hayano}, Y., {Henning}, T.,
  {Hodapp}, K.~W., {Kandori}, R., {Klahr}, H., {Kudo}, T., {Kusakabe}, N.,
  {Mordasini}, C., {Morino}, J., {Suto}, H., {Suzuki}, R., and {Tamura}, M.
  (2009) ,
\newblock {\em \apjl} {\bf 707}, L123

\bibitem[\protect\astroncite{{Todorov} et~al.}{2010}]{todorov_10}
{Todorov}, K., {Luhman}, K.~L., and {McLeod}, K.~K. (2010) ,
\newblock {\em \apjl} {\bf 714}, L84

\bibitem[\protect\astroncite{{Torres} et~al.}{2008}]{torres_08}
{Torres}, C.~A.~O., {Quast}, G.~R., {Melo}, C.~H.~F., and {Sterzik}, M.~F.
  (2008) ,
\newblock {\em {Young Nearby Loose Associations}}, pp. 757--+

\bibitem[\protect\astroncite{{Trauger} and {Traub}}{2007}]{trauger_traub_07}
{Trauger}, J.~T. and {Traub}, W.~A. (2007) ,
\newblock {\em \nat} {\bf 446}, 771

\bibitem[\protect\astroncite{{Uchimoto} et~al.}{2008}]{uchimoto_08}
{Uchimoto}, Y.~K., {Suzuki}, R., {Tokoku}, C., {Ichikawa}, T., {Konishi}, M.,
  {Yoshikawa}, T., {Omata}, K., {Nishimura}, T., {Yamada}, T., {Tanaka}, I.,
  {Kajisawa}, M., {Akiyama}, M., {Matsuda}, Y., {Yamauchi}, R., and
  {Hayashino}, T. (2008) ,
\newblock {\em \pasj} {\bf 60}, 683

\bibitem[\protect\astroncite{{Veras} et~al.}{2009}]{veras_09}
{Veras}, D., {Crepp}, J.~R., and {Ford}, E.~B. (2009) ,
\newblock {\em \apj} {\bf 696}, 1600

\bibitem[\protect\astroncite{{Wahhaj} et~al.}{2011}]{wahhaj_11}
{Wahhaj}, Z., {Liu}, M.~C., {Biller}, B.~A., {Clarke}, F., {Nielsen}, E.~L.,
  {Close}, L.~M., {Hayward}, T.~L., {Mamajek}, E.~E., {Cushing}, M., {Dupuy},
  T., {Tecza}, M., {Thatte}, N., {Chun}, M., {Ftaclas}, C., {Hartung}, M.,
  {Reid}, I.~N., {Shkolnik}, E.~L., {Alencar}, S.~H.~P., {Artymowicz}, P.,
  {Boss}, A., {de Gouveia Dal Pino}, E., {Gregorio-Hetem}, J., {Ida}, S.,
  {Kuchner}, M., {Lin}, D.~N.~C., and {Toomey}, D.~W. (2011) ,
\newblock {\em ArXiv e-prints}

\bibitem[\protect\astroncite{{Wallace} et~al.}{2010}]{wallace_10}
{Wallace}, J.~K., {Burruss}, R.~S., {Bartos}, R.~D., {Trinh}, T.~Q., {Pueyo},
  L.~A., {Fregoso}, S.~F., {Angione}, J.~R., and {Shelton}, J.~C. (2010) ,
\newblock In {\em Society of Photo-Optical Instrumentation Engineers (SPIE)
  Conference Series}, Vol. 7736 of {\em Presented at the Society of
  Photo-Optical Instrumentation Engineers (SPIE) Conference}

\bibitem[\protect\astroncite{{Watson} et~al.}{2010}]{watson_10}
{Watson}, C.~A., {Littlefair}, S.~P., {Collier Cameron}, A., {Dhillon}, V.~S.,
  and {Simpson}, E.~K. (2010) ,
\newblock {\em ArXiv e-prints}

\bibitem[\protect\astroncite{{Weinberg} et~al.}{1987}]{weinberg_87}
{Weinberg}, M.~D., {Shapiro}, S.~L., and {Wasserman}, I. (1987) ,
\newblock {\em \apj} {\bf 312}, 367

\bibitem[\protect\astroncite{{Wizinowich} and {Campbell}}{2009}]{wizinowich_10}
{Wizinowich}, P. and {Campbell}, R. (2009) ,
\newblock In {\em American Astronomical Society Meeting Abstracts}, Vol. 214 of
  {\em American Astronomical Society Meeting Abstracts}, p. 219.01

\bibitem[\protect\astroncite{{Wright} et~al.}{2009}]{wright_09}
{Wright}, J.~T., {Upadhyay}, S., {Marcy}, G.~W., {Fischer}, D.~A., {Ford},
  E.~B., and {Johnson}, J.~A. (2009) ,
\newblock {\em \apj} {\bf 693}, 1084

\bibitem[\protect\astroncite{{Zuckerman} and {Song}}{2004}]{zuckerman_song_04}
{Zuckerman}, B. and {Song}, I. (2004) ,
\newblock {\em \araa} {\bf 42}, 685

\end{thebibliography}
\end{small}

\end{document}